\documentclass[twocolumn,preprintnumbers, amsmath, amssymb]{revtex4}
\def\be{\begin{equation}}
\def\ee{\end{equation}}
\def\beq{\begin{eqnarray}}
\def\eeq{\end{eqnarray}}
\def\n{\nonumber}
\def\bay{\begin{array}}
\def\eay{\end{array}}
\def\kakueq{{\raisebox{-2.0ex}{$\sim$} \atop \raisebox{0.8ex}{$_{\mathrm{K}}$}}}

\begin{document}
%%%%%%%%%%%%%%%%%%%%%%%%%%%%%%%%%%%%%%%%%%%
\preprint{CIRI/04-smw01}
\title{ Some fundamental issues in General Relativity and their resolution}

\author{Sanjay M. Wagh}
\affiliation{Central India Research Institute, \\ Post Box 606,
Laxminagar, Nagpur 440 022, India\\
E-mail:cirinag\underline{\phantom{n}}ngp@sancharnet.in}

\date{February 2, 2004}
\begin{abstract}
The purpose of this article is to draw attention to some
fundamental issues in General Relativity. It is argued that these
deep issues cannot be resolved within the standard approach to
general relativity that considers {\em every\/} solution of
Einstein's field equations to be of relevance to some,
hypothetical or not, physical situation. Hence, to resolve the
considered problems of the standard approach to general
relativity, one must go beyond it. A possible approach, a theory
of everything, is outlined in the present article and will be
developed in details subsequently.
\\

\centerline{Submitted to:}
\end{abstract}
\maketitle

\newpage

\section{Introduction} Historically, Newton based his laws of
mechanics on the notion of a particle - a mass-point. Since then,
the notion of a mass-point has been an exceptionally useful and
important approximation in Physics, in general.

As is well known, in the Newtonian theoretical framework, a
physical body is a collection of mass-points. Newtonian laws of
motion, together with additional assumptions about the nature of
the body, such as its rigidity etc., determine the dynamics of
that body under scrutiny. It may be emphasized that these
additional assumptions are mainly ad-hoc in nature, but are very
useful nonetheless. Further, these ad-hoc assumptions primarily
deal with the nature of non-gravitational phenomena affecting the
constituent particles of a body. Newtonian inverse-square law of
gravitation, Newtonian laws of motion and these additional
assumptions then provide the basis for Newton's ``mechanical''
world-view.

Any radical departure from the Newtonian concept of a
point-particle was not evident in Physics in any concrete form
until the advent of electromagnetism in its classically complete
form. This departure gradually emerged in the field conception of
Maxwell and Faraday. With Maxwell and Faraday, the electromagnetic
field emerged as a physical entity completely separate in
existence from a Newtonian particle. Essentially, a particle, as a
conception, has ``existence'' only at one spatial location while
the field, as a conception, has ``existence'' at more than one,
continuous, spatial locations at any instant of time.

However, a point-particle is still needed in the form of a source
of the electromagnetic field. In this dualism of a field and a
particle, a source particle is necessarily a singularity of the
field it generates. For example, an electric charge, as a source,
is a {\em singularity\/} of {\em its\/} electric field. As another
example, a material point at rest is represented by its
gravitational field which is everywhere finite and regular, except
at the position where a material point is located: there the
gravitational field also has a {\em singularity}.

In this dualism, the ``gravitational'' mass and an electric charge
are the physical attributes of a particle as a result of which it
{\em generates\/} its gravitational and electric fields. In a
similar spirit, a particle could be endowed with other physical
attributes such as spin, for example.

As a separate remark, it may be also noted here that, within the
field conception, the Newtonian concept of instantaneous action at
a distance gets, naturally, replaced with that of an interaction
mediated by the field carrying the disturbance to a different
location.

Now, let us, for the moment, assume that space and time do not
play any ``dynamic" role in physical phenomena but provide only an
inert stage for them. This, for example, has been the situation
with all the pre-general relativistic theories of the physical
phenomena.

Then, a physical body is imagined to be a collection of particles
acting as sources (or singularities) of different fields depending
on the attributes chosen for the particles making up the body.

Now, we could approximate any collection of mass-points by average
mass-density. This is permissible since a volume form, in Cartan's
sense, of the background space is available to define a smooth
distribution of sources. (This will be commented on at a later
stage.)

This is the {\em fluid description\/} of sources. As is
well-known, a continuous distribution allows fields to be defined
at points outside as well as ``inside'' the distribution of
sources itself since the field singularities have been done away
with.

Now, since the volume-form of the space is well-defined at all
points of the background space, well-posed problems could be
formulated for which some fluid distribution of sources generated
a ``smooth" field even when the field is singular at the locations
of its sources.

Similarly, some smooth distribution of sources could also be
replaced by a ``concentrated" source, a very useful approximation
in Physics. %\cite{geroch-traschen}.
For example, a charge distribution could be replaced by a point
charge or a mass distribution could be replaced by a point mass.

In either case, it can be uniquely ascertained as to when the
fluid description fails for a given situation. The inert stage of
space and time provides the means of these assertions.
Essentially, the fluid description fails when the volume of
interest is so small that fluctuations in variables due to
``individual'' particles become significant.

\section{Situation in General Relativity}

General Theory of Relativity (GTR) changes the canvas of this
painting drastically. A spacetime becomes a ``dynamic" stage for
physical phenomena. The stage changes with the events and events
in turn change as the stage changes.

Firstly, a point-particle in GTR is, necessarily,  a {\em
curvature singularity\/} of the spacetime.

A point-particle has existence only at one spatial location and
everywhere else, except at this {\em special\/} location, it is
the non-singular gravitational field of the particle that
``exists''. In its spherically symmetric spacetime {\em geometry},
all the points, except for the location of the mass-point, are
then ``matter-wise equivalent'' to each other as the rest of the
manifold describes the vacuum ``gravitational field'' that only
diverges at the location of the mass-point unless, perhaps, we
include the self-field in some way. This {\em geometry\/} then
should not ``include'' the location of the mass-point. Then, any
mass-point is a curvature singularity of the manifold describing
its ``exterior'' gravitational field.

Addition of other attributes of a particle, like charge, spin
etc., does not change this situation. The Kerr-Newman family
exists in GTR and the point-mass with charge and spin is a
spacetime singularity in it. (The issue of the inadequacy of such
a description of a physical particle would have arisen if we had
no Kerr-Newman family of spacetimes in GTR. We could then have
said that the pure mass-point is an inadequate description of a
``physical particle'' and the addition of other attributes ``takes
us away from General Relativity'' to some new theory with some new
description of a physical particle.)

If we base our picture of the physical world on the notion of a
point-particle then, a physical body will be a collection of
point-particles, a collection of spacetime singularities. As was
the case with Newton's mechanical world-view, we will have to {\em
assume\/} that non-gravitational interactions of these
point-particles will be responsible for the stability, rigidity
etc.\ of physical bodies.

Now, if there existed other particles then, the spacetime geometry
would be {\em different\/} from that of a single point-particle
and the geodesics of that geometry would not be those of the
geometry of a single point-particle.

Furthermore, non-gravitational interactions between various
particles will have to be incorporated in the geometry, as
attributes of a particle. This is also a relevant issue.

Strictly speaking, a path of a point-particle is not a geodesic
since the mathematical structure defining a geodesic breaks down
at every point along the path of the singularity.

We may surround the singular trajectory of a point-particle by an
appropriately ``small'' world-tube, remove the singular trajectory
and call this tube the ``geodesic'' of the particle. But, the new
spacetime is {\em not\/} the original spacetime and, by the
physical basis of GTR, it corresponds to different gravitational
source. Therefore, this procedure is not at all satisfactory.

We may, along a geodesic, change the spatial coordinate label(s),
by which we ``identify the singularity'', with respect to the time
label of the spacetime geometry and may term this change the
``motion'' of a particle along a geodesic. However, this is only
some ``special'' simultaneous relabelling of the coordinates. But,
a particle in the spacetime does not ``move'' at all.

The motion of a particle must be a ``singular trajectory'' in the
spacetime if a particle, the singularity, ``moves'' at all. It is
also equally clear that any singular trajectory is {\em not\/} a
geodesic in the spacetime since such a trajectory is {\em not\/}
the part of the smooth spacetime geometry.

Then, strictly speaking, geodesics of a spacetime do {\em not\/}
provide the {\em law of motion\/} of particles. Thus, strictly
speaking, we have to specify {\em separately\/} the law of motion
for the particles. And, GTR does not, strictly speaking, specify
this law of motion for the particles. This peculiar situation is
quite similar to that of the Newtonian theoretical framework then.

Ignoring these difficulties, we may now replace a collection of
point-particles by that of their smooth distribution. But, a new
difficulty, not encountered in Newton's theory, arises as
particles are spacetime singularities.

Essentially, a spacetime geometry with ``curvature singularities''
is being replaced by a smooth geometry using the volume-form of
the geometry with curvature singularities. Therefore, such a
construction has its own pitfalls.

A point-particle is an essential, non-removable, singularity of
the spacetime geometry. Thus, approximating a geometry with a
singularity by any smooth manifold means that an entirely
different source of gravity is acting in the latter situation.
This is, physically, not acceptable.

Also, replacing a collection of point-particles by a smooth
manifold is mathematically unacceptable for the following reasons.

Let us attach a ``weight function'' to each mass-point in our
collection. Then, the weight-function must ``diverge at suitable
rate'' at the mass-point to balance the field-singularity at its
location. The resulting ``weighted-sum'' should provide a smooth
``volume-measure'' as well as a smooth density with respect to the
volume-measure of the geometry (that is being replaced).

In terms of the differentials of measures, $d\mu = \rho(x)
\,d\mu_o$ where $\mu$ is the volume-measure of the smooth
geometry, $\mu_o$ that of the geometry with singularities and
$\rho(x)$ is the required density. Any smooth $\rho(x)$ is {\em
impossible\/} if the geometry of $\mu_o$ has essential curvature
singularities. To keep $\mu$ smooth the density $\rho(x)$ must
diverge at the curvature-singularities of $\mu_o$.

In the newtonian situation, the geometry of $\mu_o$-measure and
the geometry of $\mu$-measure, both, are well-behaved. It is {\em
not\/} the background spacetime which is singular but only some
field defined over it that is singular. If we now select some
appropriate smooth $\rho(x)$, in the above construction, it can
``replace the singularities'' of the field by means of suitable
weight-functions correcting the field-singularity at locations of
mass-points. Therefore, there are no fundamental difficulties of
any kind with this procedure in the newtonian theory that deals
only with fields defined over a smooth, flat, spacetime manifold.

Moreover, consider that a smooth spherical spacetime has ``mass
function'' $m=4\pi \int \rho r^2 dr$ where $\rho(r)$ is the {\em
smooth\/} energy density that is non-vanishing in some region
around $r=0$ and $r$ is suitably defined radial coordinate. As we
approach the point $r=0$, $m\to 0$ since $\rho$ is a non-vanishing
function of $r$ in that region.

But, the fluid approximation of sources must fail as the ``volume
of interest'' is so small that fluctuations in density due to
``individual'' particles become significant much before we
approach the point $r=0$. Then, in a spherically symmetric
distribution, there is a mass-point exactly at $r=0$ or there is
none. Both these possibilities are certainly allowed, even
generally. Then, $m \to 0$ or $m\nrightarrow 0$ at a location
inside the fluid.

In the newtonian case, we could avoid these problems by appealing
to the well-behaved volume form of the inert background spacetime
and by treating the particle as of arbitrarily small volume and of
correspondingly high density. Moreover, as coordinate differences
correspond to physical distances in this theory, we could always
impose $m \to 0$ or $m\nrightarrow 0$ on the newtonian solutions
in an equivalent manner.

But, this is not the situation with GTR. A particle is an {\em
essential singularity\/} of the spacetime structure and no smooth
spacetime can {\em replace\/} this essential singularity without
changing the underlying physical basis of this theory.

Essentially, fluid approximation breaks down and there is no
mathematical procedure to circumvent associated problems in the
absence of ``background'' spacetime geometry to appeal to.

Furthermore, coordinate differences are not the geometrical or
physical distances in this theory. If the density $\rho$ is
non-vanishing inside the source distribution then, for $m\to 0$,
we have $r\to 0$ but, for $m\nrightarrow 0$, we have,
$r\nrightarrow 0$. These two conditions are then the {\em
coordinate freedom\/} in GTR.

But, it is customary to impose only the first of these two
conditions while discarding the second as being unphysical. That
is to say, it is customary to demand that the spacetime locations
be {\em regular\/} in the sense that the orbits of the $SO_3$
group of rotations shrink to a zero radius ($r\to 0$) at every
such location thereby discarding the situation of $r\nrightarrow
0$ as being unphysical. But, as the particle picture shows it, the
condition that $r\nrightarrow 0$ is also {\em physically\/}
motivated and allowed one.

Now, let us keep these difficulties aside and let us, to fix
ideas, consider a spherical star using the fluid approximation.
Then, we have some smooth spherical spacetime for this star.

Let us consider two copies of such a star separated by a very
large distance today so that the evolution of one cannot
essentially affect the evolution of the other. Evolution of the
two copies may be expected, on the basis of general physical
considerations, to be identical.

In the newtonian theory, this situation can be treated as a
two-body problem with evolution of each star being independent of
the other except for the, extremely weak, gravitational
interaction of the two stars. The outcome of the collapse of each
star is therefore identical here.

In General Relativity, however, the situation is more complicated
than this. The main reason for the complication is that the
spacetime of the two stars taken together,  no matter how far away
the two stars are, is {\em not\/} globally spherically symmetric
even though the spacetime of individual stars is globally
spherically symmetric. Let us also ignore this problem, since our
assumption, that the evolution of each star will be ``weakly''
affected by the other distant star, is still valid.

Let both these stars begin to collapse. Let us focus attention on
one star. Then, for the newtonian and general relativistic
situations, both, the presence of a very distant star should
``weakly'' affect the evolution of that star. So, if for our
globally spherical spacetime, we had some definite outcome in the
collapse of the star, there will be exactly the same outcome
observable at locations of each star in the above situation.

We may add further copies of the same star, again separated by the
{\em same\/} large distance from original stars and from each
other. If every star collapsed to a definite outcome, there will
be the same result observable at locations of each star in this,
completely arbitrary, situation too.

Let us list our assumptions separately: \begin{itemize} \item the
presence of a very distant star ``weakly'' affects the evolution
of any particular star at a given epoch,
\item the collapse of an individual star, for the spacetime of an
individual star, results in some definite outcome, say, as a naked
singularity or a black hole.
\end{itemize} Then, if we have just one example of the
gravitational collapse leading to a naked singularity for the
spherical spacetime then, the Cosmic Censorship, that spacetime
singularities are not visible to any observer, is certainly
invalid!

Is this conclusion of the non-validity of the cosmic censorship
really justified? That is to say, do we have faith in the above
arguments in either the newtonian theory or in GTR?

In the newtonian theoretical framework, we have no reason to doubt
this conclusion. Each star collapses to a point and we have the
newtonian concept of a point-particle to back it up. We then know
how to treat a point-particle in the newtonian theory. We then do
not need any Cosmic Censor in Newton's theory.

Now, let us keep aside the question of the validity of the fluid
approximation as we approach a spacetime singularity. Then, in GTR
too, each spherical star collapses to a point-particle, that is, a
spacetime singularity.

Now, we may find that in some situations a {\em null\/} trajectory
emanates from this singularity, that is, the singularity is naked.
Or, in some other situations, we may find that the singularity is
covered by a {\em horizon}, that is, there is a black hole. That
is, equivalently, a point-particle is naked or is covered by a
horizon in GTR.

Then, since we have accepted a point-particle in GTR, we may
renounce the Cosmic Censorship even in GTR and may accept the
existence of naked singularities and black holes, both, as being
physically or astrophysically relevant. Then, we choose to ignore
all the serious problems of GTR as have been discussed earlier.

But, the situation with GTR is definitely not so simple and
straightforward as it appears from the above discussion. The main
reason is that a spacetime necessarily has {\em information\/}
about the ``past'' and ``future'', both. That the two stars are
separated by very large distance {\em today\/} does {\em not\/}
mean that they were always so separated.

It could, for example, be that the two stars were ``close'' in the
past but due to the directions of their respective velocities they
moved away from each other to be separated by large distance at
the present epoch.

What is {\em important\/} is that it is the {\em same spacetime\/}
that is prescribed by GTR in either of these situations, of stars
separated by some ``small'' distance or of stars separated by some
very ``large'' distance between them. General Theory of
Relativity, as a theory of gravitation, has no length-scale of any
kind to distinguish, in any manner, between these two situations.

Now, let us return to the newtonian situation to consider an
important related issue.

Notice then that the newtonian theory has {\em linear\/} laws.
Such linear laws obey the superposition principle for their
solutions. In any linear theory, particles must therefore be
provided with {\em independent attributes\/} for any interactions
and the corresponding laws for such interactions will also have to
be mutually independent. Therefore, such linear laws cannot
intrinsically contain any assertions about the interactions of
elementary bodies, the particles. Hence, if we aim for the
unification of physical laws, the theory cannot be linear nor can
it be derived from such linear laws.

Furthermore, the linearity of the newtonian laws is also the
primary reason as to why we can ``superimpose'' the individual
newtonian solutions of each star to obtain the solution for the
situation of combined stars. This is also the primary reason as to
why we can, equally well, treat the ``weak'' gravitational effects
of the presence of another very distant star as a perturbation of
the newtonian solution of the individual star.

But, the spacetime of combined stars is not obtainable as a
superposition of two or more number of spherical spacetimes. GTR
is an intrinsically nonlinear theory of gravitation. The solutions
of the highly nonlinear field equations of GTR do not follow the
superposition principle, in any conceivable manner whatsoever.

Then, the ``perturbations of spherical spacetime'' cannot,
fundamentally, be faithful to the physical situation of combined
stars. The spacetime of combined stars may be approximated by the
perturbations of the spherical spacetime of a single star, but the
corresponding results of physical significance cannot be
faithfully obtained from them as was the case with a linear
theory. Needless to say here then, the ``perturbations of
spacetime'' obey linear laws. %12345

This peculiar situation in GTR means that the outcome of the
gravitational collapse of an individual star in a spacetime of the
individual star is {\em not necessarily obtainable\/} in the
spacetime of combined stars. Without adequate investigations, it,
therefore, prevents us from accepting the earlier arguments
leading to non-validity of the Cosmic Censorship in GTR.

Moreover, ``some properties'' of the spacetime of an individual
star will not necessarily be those of the spacetime of stars taken
together. That is to say, non-linearity of the field equations of
GTR does not permit us to conclude, without adequate justification
or investigation, that the spacetime of combined stars has some
properties similar to the spacetime of an individual star.

Hence, without adequate justification or investigation, we cannot
conclude that any specific property, in particular, the presence
of an event horizon, of the spacetime of an individual star is
obtainable quite generally for the spacetime of combined stars.
That is to say, we cannot similarly conclude that even black holes
will form in the spacetime of combined stars.

General Theory of Relativity is an intrinsically non-linear theory
of gravitation.

\section{Field-particle dualism in General Relativity} Let us
now, in the light of all the earlier issues, reconsider the
field-particle dualism in General Theory of Relativity.

Firstly, in these discussions, GTR is being considered as a theory
of the ``pure'' gravitational field, other fields of nature not
being ``incorporated'' in the (Schwarzschild) geometry of a
point-mass, it has only the ``gravitational mass'' attributed to
it. That is, we consider that this (Schwarzschild) geometry
describes only the ``pure'' gravitational field of a mass-point.

Then, since a physical particle has other attributes, its
spacetime geometry should the less be viewed as a pure
gravitational field, as described by the Schwarzschild geometry,
the closer one comes to the position of a particle - that is again
a spacetime singularity.

Now, the concept of a ``pure'' gravitational field is a
reminiscent of the Newtonian concept of the gravitational field of
a point particle.

In Newton's theory, the ``gravitational mass'' of a particle is
just one among the possible attributes of a particle. Each
attribute of a particle produces the corresponding field. In this
linear theory, any other field attribute of a particle cannot be
related to or be determined by its gravitational mass attribute.
It's laws are linear laws and, hence, we have to separately
postulate the laws of interactions among the particles. Separate
fields obey separate, meaning mutually independent, laws in this
theoretical framework.

That is why we can speak of a ``pure'' gravitational field and
``other'' fields in a linear theory as is Newton's theory. Mutual
independency of the laws for different fields makes this possible.
Notice that it is mutual independency that is crucial in this
theoretical framework.

But, in GTR, every attribute of a particle affects its spacetime
geometry. That is, associated with every attribute of a particle
corresponding to any field of Nature, there must be an appropriate
corresponding ``effect'' on the spacetime geometry of the
particle. This is inevitable in an intrinsically non-linear theory
as is GTR. [The well-known examples of Reissner-Nordstrom and
Kerr-Newman spacetimes clearly show this to us.]

Clearly, ascribing attributes, other than gravitational mass, to a
particle in GTR is then equivalent to changing the spacetime
geometry of a point mass that ``continues'' to be a non-removable,
essential, spacetime singularity.

That is to say, a particle or a mass-point remains a spacetime
singularity even after we account for, let us say, all the
possible attributes of other fields of Nature in this way.

Then, many fundamental problems associated with the existence of a
spacetime singularity, that have been discussed earlier, are
unavoidable for this field-particle dualism.

Then, strictly speaking, we must formulate ``external'' laws for
interactions of particles. Further, the attributes of a particle,
such as its mass or charge, are {\em additive\/} and, hence, the
required ``external'' laws of interactions of particles will,
necessarily, be {\em linear or additive\/} in all these attributes
of a point-particle.

But, all this is certainly contrary to the physical basis of GTR!
In essence, a point-particle is inherently inconsistent with GTR.

Therefore, in view of the fundamental problems discussed earlier,
it is then essential to renounce the field-particle dualism in
GTR. That is to say, we renounce the picture of a point-particle,
a spacetime singularity in GTR.

It may be noticed now that any open or concealed increasing of the
number of dimensions from four does not improve the above
problematic situation with the field-particle dualism.

Again, a point-particle will be a curvature singularity of the
higher dimensional manifold. Then, if we continue with the
field-particle dualism, all the fundamental problems associated
with the singularity of the higher-dimensional spacetime manifold
will continue to haunt us in the same manner as holds for the
four-dimensional case of General Relativity. It then seems highly
unlikely that any higher dimensional theory will, while continuing
with the field-particle dualism, be able to provide any ``basis''
for the physical phenomena than that already provided by GTR.

Thus, we then consider here only singularity-free, smooth,
four-dimensional spacetime geometries in General Relativity. Many
such spacetime geometries are obtainable in GTR.

Then, an important question is: which of these smooth spacetime
geometries are to be considered physically relevant?

Before we consider this issue, it is instructive to consider, on
the basis of very general physical considerations, the physical
properties that any such spacetime geometry should possess.

Firstly, we note that, in a spherically symmetric such spacetime,
we will necessarily have $m=4\pi\int \rho r^2 dr \nrightarrow 0$
at the center of the spacetime since $\rho$ is a non-vanishing
function of the radial coordinate and the well-behaved volume-form
of any such smooth spacetime allows us to ``define'' an
equivalent, non-vanishing, ``gravitational'' mass at its central
spatial location.

That is to say, in such a smooth spacetime geometry, we could
always ``treat a gravitational mass'' as of arbitrarily small
volume and of correspondingly high density as a useful
approximation. Then, for such a spherically symmetric spacetime,
we will have $m\nrightarrow 0$ and, hence, $r\nrightarrow 0$. This
will be the situation even in a general, non-symmetric, such
spacetime.

Secondly, we note that any physical object is to be treated as a
``spatially'' concentrated form of {\em energy\/} in such a smooth
spacetime geometry. As this object moves in space, the
energy-distribution changes in the space. Then, the ``motion'' of
any object is ``equivalent'' to change in the energy-distribution
in such a spacetime. As we ``move'' any given object in the space,
the energy-distribution changes accordingly and, hence,
energy-distribution should be arbitrary if a physical body moves
so in such a spacetime.

Remarkably, this picture is also consistent with the following
observation that can be said to form the basis for cosmology.

Gravitational and other fields have been rearranged on the Earth
on many instances. That is to say, the local distributions of
different fields of Nature have been (and can be) altered on the
Earth in an {\em arbitrary\/} manner at any time.

Provided that the Earth is not any special location in the
universe, the {\em Copernican principle}, the local source
distribution of different fields of Nature can be altered in an
{\em arbitrary\/} manner {\em anywhere and at any time \/} in the
universe. This would be possible if the distribution of fields
were arbitrary in a spacetime continuum.

Clearly, the above two entirely different approaches reach the
same conclusion in the form of the {\em required spacetime}. This
cannot be just a mere simple coincidence.

As a matter of principles, we should then not consider an {\em
isolated object, or its spacetime, or the outcome of its
collapse}, as being physically relevant since no object is
genuinely isolated in the universe. The non-linearity of the field
equations of GTR then clearly indicates that such spacetime
geometries will get {\em modified\/} when we embed the
corresponding objects in the universe. Properties of the modified
spacetime geometry will not necessarily be those of the spacetime
of an isolated object. Thus, the spacetime of an isolated physical
object cannot be a physically relevant or meaningful spacetime
geometry.

Therefore, the issue of cosmic censorship for such spacetimes of
isolated or individual physical objects like an isolated star is
not any relevant issue for the physical theories.

Furthermore, as a matter of principles again, we should similarly
not consider any smooth, singularity-free spacetime geometry that
does not,  without changing its global mathematical properties,
let us ``shift physical objects'' arbitrarily in it to be
physically relevant or meaningful. We do shift ourselves
arbitrarily in the ``spacetime geometry of the universe'' we live
in.

The non-linearity of the field equations of GTR then clearly
indicates in this case that such spacetimes will get {\em modified
globally \/} when we ``shift'' physical objects in them. Again,
the nonlinearity of the field equations of General Relativity then
indicates that the global properties of the modified spacetime
will not necessarily be those of the original spacetime.

Therefore, if in a chosen spacetime of the universe, we cannot
``shift'' physical objects {\em arbitrarily\/} in a local spatial
region without changing global spacetime properties, it is
difficult to imagine as to how that spacetime can faithfully
represent the {\em observable universe}.

One now sees clearly the emerging confluence of the above two
approaches. The two approaches demand the existence of the {\em
same}, mathematically smooth, spacetime geometry.

{\em Therefore, the geometric properties of an appropriate smooth,
singularity-free spacetime, one with arbitrary spatial properties,
are the only that are genuinely physically relevant ones. This is,
of course, subject only to the caveat that the theoretical
description of physical systems based on any geometric continuum
is permissible}.

As this spacetime is {\em singularity-free\/} everywhere, its
initial data is singularity-free or {\em regular\/} everywhere on
its spacelike hyper-surfaces. The question of cosmic censorship
then reduces to whether any observable spacetime singularity
results in temporal evolution of this initial data. The issue of
cosmic censorship, if any, can then be settled only when we have,
explicitly, such a spacetime geometry to study.

Hence, only if the naked singularities or event horizons arise for
such a smooth spacetime, can these be physically relevant
concepts.

\section{Beyond standard formalism of General Relativity}

We thus consider here only smooth spacetime geometries in GTR.
Moreover, any geometric theory of gravitation, as is GTR, must
then incorporate all the permissible field-attributes of various
possible fields of Nature in some mathematically smooth spacetime
geometry. Thus, there must then exist a spacetime incorporating
all the possible fields of Nature in it. After all, the fields are
continuous, live in space and evolve in time.

There must therefore exist a spacetime in such a theory that has
all the ``fields'' of Nature incorporated in its geometry. Such a
spacetime can be considered to provide the geometry of the {\em
total field}, in Einstein's sense \cite{ein1}.

Consequently, some general relativistic spacetime must then be a
{\em complete spacetime geometry\/} in which ``particles''
themselves, as concentrated form of energy but not as spacetime
singularities, would {\em everywhere in space\/} be describable as
singularity-free.

Provided, of course, that the theoretical description of
``physical systems'' is permissible on the basis of a geometric
spacetime continuum, as is the underlying basis of the General
Theory of Relativity, this is then the logically inevitable
conclusion. Such a spacetime then admits ``arbitrary''
distribution of fields in it.

The required spacetime that is everywhere regular and
singularity-free has a (pseudo-)metric of the form:
\beq ds^2 &=& -P^2Q^2R^2dt^2  +\;\gamma^2{P'}^2Q^2R^2B^2 dx^2 \n \\
&\phantom{=}& \qquad +\;\gamma^2P^2\bar{Q}^2 R^2C^2dy^2 \n \\
&\phantom{=}& \qquad +\;\gamma^2P^2Q^2\tilde{R}^2D^2\,dz^2
\label{genhsp} \eeq where $P\equiv P(x)$, $Q\equiv Q(y)$, $R\equiv
R(z)$, $B\equiv B(t)$, $C\equiv C(t)$, $D\equiv D(t)$ and $\gamma$
is a constant. Also, $P'=dP/dx$, $\bar{Q}=dQ/dy$ and
$\tilde{R}=dR/dz$.

Now, as can also be checked easily, the metric (\ref{genhsp})
admits, precisely, three spacelike homothetic Killing vectors.
From Lie's theory of differential equations, it then follows that
there will be three {\em arbitrary\/} functions of spatial
coordinates corresponding to these three spacelike homothetic
Killing vectors. These are the spatial functions $P$, $Q$, $R$. We
also note here that the spacetime of (\ref{genhsp}) is also a
machian spacetime \cite{mach, smw1, smw2}.

In general, there are two types of curvature singularities of the
metric (\ref{genhsp}). Singularities of the first type, vanishing
of any temporal functions $B$, $C$, $D$, are singular
hyper-surfaces while singularities of the second type, when any
one of the spatial functions $P$, $Q$, $R$ is vanishing, are
singular spatial data for (\ref{genhsp}).

The locations for which the spatial derivatives vanish are,
however, coordinate singularities. The curvature invariants of
(\ref{genhsp}) do not blow up at such locations. There are also
obvious degenerate metric situations when any of the spatial
functions is infinite for some range of the coordinates.

We may now consider an appropriate energy-momentum tensor for the
fluid in the spacetime and write down the field equations of GTR.
But, as can be verified, the field equations do not determine the
spatial functions $P$, $Q$, $R$.

The energy-density in the spacetime of (\ref{genhsp}) varies as
$\rho \propto {1}/{P^2Q^2R^2}$ and is {\em arbitrary\/} since the
field equations do not determine the spatial functions $P$, $Q$,
$R$.

But, it is known \cite{subtle} that the energy-momentum tensor is
not a satisfactory concept in General Relativity \footnote{As
Einstein \cite{ein1} says it: ``The right side (of
$R_{ik}-\frac{1}{2}g_{ik}R=-\,k\,T_{ik}$: my brackets) is a formal
condensation of all things whose comprehension in the sense of a
field-theory is still problematic.'' He is referring to the
comprehension of the energy-momentum tensor in terms of
``extended'' bodies (fields) and not point-particles. That he
recognizes the dependence of the usual conception of the
energy-momentum tensor on that of a point-particle is clear.}.
Reason for this appears to be its dependence on the concept of a
point-particle.

Recall that to obtain the energy-momentum tensor we consider a
collection of particles and try to obtain a suitable {\em
continuum approximation\/} of different physical quantities such
as energy density, momentum flux etc.

In the newtonian situation this procedure works well because the
background spacetime geometry is non-singular.

But, in the standard approach to GTR, wherein we consider each
spacetime geometry as a solution of the Einstein field equations
to be of relevance to some, hypothetical or not, physical
situation, the spacetime singularities do not permit us any
meaningful averaging procedure, as was seen earlier. Consequently,
various difficulties in defining the energy-momentum tensor arise
in this situation and doom the use of the energy-momentum tensor
to be a failure.

Thus, the problem, now, is that the energy-momentum tensor is, in
general, not a well-defined concept. Therefore, we are not
justified in using it to obtain the Einstein field equations.
Then, we do not ``know'' for sure as to how to determine the three
temporal functions in the metric (\ref{genhsp}) completely. We
therefore have no tools at hands to construct the four-dimensional
spacetime geometry of the metric (\ref{genhsp}).

That is why, we essentially abandon the four-dimensional picture
and consider only a smooth, three-dimensional manifold admitting a
positive-definite metric of the form (\ref{3d-metric-gen}). (See
below.)

What is now so crucial to realize is that there are three temporal
functions $B$, $C$, $D$ in the metric corresponding to ``motion''
along three spatial directions. Consequently, we may then consider
a physical object (on any $t=$ constant hyper-surface of this
spacetime) and consider that it ``moves'' with velocity having
appropriate components along the three spatial directions.

Thus emerges the picture of ``motion'' in this spacetime that
motion of a physical object corresponds only to a change in the
energy density in this spacetime.

Therefore, we essentially abandon the four-dimensional spacetime
as a basic picture and consider only a smooth, three-dimensional
manifold admitting a positive-definite metric of the form
\beq d\ell^2&=& {P'}^2Q^2R^2\, dx^2 \n \\
&\phantom{m}&\hspace{.1in} +\;P^2\bar{Q}^2 R^2\, dy^2 \n \\
&\phantom{m}&\hspace{.3in} +\;P^2Q^2\tilde{R}^2 \,dz^2
\label{3d-metric-gen} \eeq where $P(x)$, $Q(y)$, $R(z)$ are
arbitrary functions of their respective arguments. We shall call
the space of the metric (\ref{3d-metric-gen}) as the {\em base
space\/} or {\em Einstein space}. It will be denoted by the symbol
$\mathbb{B}$.

Any specific choice of functions, say, $P_o$, $Q_o$, $R_o$ gives
us a specific spatial distribution of energy $\rho$ in the space
of (\ref{3d-metric-gen}). As ``concentrated'' energy ``moves'' in
the space, we have the original set of functions changing to a
``new'' set of corresponding functions, say, $P_1$, $Q_1$, $R_1$.

Spatial functions $P$, $Q$, $R$ constitute {\em initial data\/}
for the base space $\mathbb{B}$. Then, ``motion'' as described
above is, basically, a {\em change of one set of initial data\/}
to {\em another set of initial data\/} with ``time''. Therefore,
we will be considering  ``motion'' as a {\em mapping from the
initial data set to the initial data set\/} of the base space
$\mathbb{B}$ \footnote{Girish Sahasrabudhe suggested this
phraseology}.

Clearly, we are considering isometries of the metric
(\ref{3d-metric-gen}) while considering ``motion'' of this kind.
That we will remain within the group of the isometries of
(\ref{3d-metric-gen}) is always ``guaranteed'' as long as we are
restricting to the triplet of functions $P$, $Q$, $R$ subject, of
course, to some other further conditions on them obtainable as
follows.

Since the vanishing of any of the spatial functions $P$, $Q$, $R$
is a {\em curvature singularity\/} of the metric
(\ref{3d-metric-gen}), we will have to restrict ourselves to
non-vanishing, strictly positive (or strictly negative),
real-valued such functions of their respective arguments. This
type of a restriction is necessary and sufficient, both, to remain
within the group of isometries of (\ref{3d-metric-gen}).

Let us then denote by ${\cal F}$ the set of all strictly positive,
nowhere-vanishing, real-valued functions from $\mathbb{R}\to
\mathbb{R}^+$. Now, consider a direct-product ${\cal F}\times
{\cal F} \times {\cal F}$ defined by an ordered triplet $\left( P,
Q, R \right) \in {\cal F}\times {\cal F} \times {\cal F}$. As can
be easily verified, this set, ${\cal F}\times {\cal F} \times
{\cal F}$, is a {\em convex\/} set.

Now, consider the set of all the possible mappings from ${\cal
F}\times {\cal F} \times {\cal F}\to \mathbb{R}$. Every such map
defines a family of ``curves'' connecting points of ${\cal
F}\times {\cal F} \times {\cal F}$ and, hence, it defines
``motion'' in our picture here. Every such map is then a {\em
suitably defined dynamical system\/} describing motion in the base
space $\mathbb{B}$. (This notion of a dynamical system will be
made mathematically precise later at a suitable stage.)

Then, different ``motions'' of any ``specific region of
concentrated energy'' are ``classifiable'' in terms of the
corresponding classification of suitably defined dynamical
systems.

Therefore, what we are looking for is a {\em complete
classification of such suitably defined dynamical systems\/}
definable on the basic product set ${\cal F}\times {\cal F} \times
{\cal F}$ or the base space $\mathbb{B}$.

\section{Mathematical Requirements}
\subsection{Mathematical short-forms} In what follows, we shall adopt
the following short-forms. We shall write for ``with respect to''
as ``wrt''. We also write ``s.t.'' for ``such that'' and ``a.e.''
for ``almost everywhere'' which means that the property holds
except for a set of measure zero. We will also write ``SBS'' for
``Standard Borel Space''. Different such short-form notations have
been listed in the Appendix.

\subsection{Formalism for dynamics}

In this connection, we firstly note that the differential of the
volume-measure on the base space $\mathbb{B}$ defined by metric
(\ref{3d-metric-gen}) is \be d\mu \;=\; P^2Q^2R^2\,\left(
\frac{dP}{dx}\frac{dQ}{dy} \frac{dR}{dz}\right)\;dx\,dy\,dz
\label{volume1} \ee This differential of the volume-measure
vanishes when any of the derivatives, of $P$, $Q$, $R$ with
respect to their arguments, vanishes. (Recall that $P$, $Q$, $R$
are non-vanishing over $\mathbb{B}$.)

We, therefore, define a set, to be called a {\bf P-set}, denoted
as $P$, by the following definition:

\begin{quotation} A {\bf P-set} of the base space $\mathbb{B}$
is the interior of a region of $\mathbb{B}$ for which the
differential of the volume-measure, (\ref{volume1}), is vanishing
on its boundary while it being non-vanishing at any of its
interior points.
\end{quotation}

Furthermore, each P-set is a metric space with a metric \[
d\ell^2\;=\;Q^2R^2dP^2 +P^2R^2dQ^2+ P^2Q^2dR^2 \] Note that all
the spatial functions $P$, $Q$, $R$ are {\em monotonic\/} within a
P-set.

Now, any two given P-sets, ${P}_i$ and ${P}_j$,
$i,j\;\in\;\mathbb{N}$, $i\,\neq\,j$, are, consequently, {\em
disjoint sets}. For, if the intersection of ${P}_i$ and ${P}_j$
were non-empty, there would be {\em interior\/} points of each of
the two P-sets at which the differential of volume-measure,
(\ref{volume1}), would be vanishing, contrary to our definition of
a P-set as given above.

Now, for a given, specific choice of, spatial functions $P$, $Q$,
$R$, {\em ie}, for a point of ${\cal F}\times {\cal F} \times
{\cal F}$ , the base space $\mathbb{B}$ is, uniquely, expressible
as a countably infinite union of the {\em closures\/} of P-sets,
{\em ie}, $\mathbb{B}=\bigcup_{i=1}^{\infty}\,({P}_i)^c$.

Furthermore, there are uncountably many such families of P-sets
with the base space $\mathbb{B}$ being equal to the countably
infinite union of the closures of the members of each such family
of P-sets. Each such family of P-sets is characterized by an
ordered triplet $(P, Q, R) \in {\cal F}\times {\cal F} \times
{\cal F}$.

Hence, ``motion'' in the present picture can also be looked upon
as a ``suitably defined map'' from one such family of P-sets to
another such family of P-sets.

A concentrated form of energy is a P-Set within this framework.
The reason for naming the set under consideration as a P-set is
then clear. A {\bf P-set} is a {\em physical particle\/} in the
present scenario. That is to say, we can associate physical
attributes of a physical particle with a P-set.

(Then, {\em attributes \/} of a physical particle will be {\em
measures\/} definable on a P-set. We will be dealing with measures
is then clear.)

Interestingly, and crucially, any P-set is {\bf\it not\/}
contractible to any of its points as a P-set. That is to say, a
singleton subset $\{\{x\}:\,x\in \mathbb{B}\}$ of $\mathbb{B}$ is
a {\em not\/} a P-set.

Each P-set is then imagined to be a {\em physical particle}. That
is to say, each P-set can be ascribed different properties of a
physical particle. Then, since a P-set cannot be shrunk to a point
as a P-set of the base space, each P-set is a physical particle
which is, fundamentally, an extended basic physical body in this
picture.

The question then arises about an appropriate mathematical
description of the dynamics of such particle systems.

As a concrete example of the current situation, let us imagine
that the 3-dimensional space (of finite or infinite total volume,
depending on the functions $P$, $Q$, $R$) is uniformly filled with
{\em solid balls\/} touching each other. There will, of course, be
{\em gap-filling spaces\/} in such a distribution which will be
having the same shape at all the corresponding locations. Then,
let us mark one solid ball as {\em Red\/} and another ``distant''
(meaning not touching the first one) as {\em Green}.

{\em We now want to mathematically describe the {\em motion\/} of
the {\em Red ball\/} so that it {\em moves towards\/} the {\em
Green ball}, touches that ball and {\em returns\/} to its original
position}. Clearly, the gap-filling spaces must change their
shapes so that the Red ball performs this motion. Moreover, even
the balls may change their shapes to move.

In the above situation, Red/Green ball is one type of a P-set and
Gap-filler space is the other type of a P-set. Any ``motion'' is
then possible iff the P-sets change their shapes and adjust
themselves as they perform the ``motion''.

Going further with this example, we may also consider that two or
more balls (some mutually touching each other) ``stick to each
other'' as they ``move collectively''. Again, some of the P-sets
must change their shapes suitably for such a motion to be possible
at all.

Therefore, a P-set may also change its {\em shape\/} during its
dynamical evolution. In fact, some P-set(s) must change their {\em
shapes\/} if at all there is to be ``motion'' in the present
picture. Then, this must also reflect in the mathematical
description being used in the present picture.

We, once again, stress that any point of the base space
$\mathbb{B}$, as a singleton set, is {\em not\/} a P-set. Hence, a
particle cannot be a point within this picture and is always an
extended body. Furthermore, a physical body is, in general, a
collection of such basic extended particles.

Hence, any such mathematical description must also be applicable
to a collection of P-sets since a physical body is imagined to be
a collection of P-sets and such a physical body ``may move'' while
maintaining its ``shape''.

In what follows, we review \cite{kdjoshi,trim6, krp, halmos1} some
mathematical aspects of relevance to the present physical
framework. Much of it is well-known (in particular, from the
ergodic theory). Nonetheless, for completeness, for fixing of
notations and for the benefit of those not familiar with the
involved terminology, we include it here.

Now, we recall that a {\em metric space\/} is a pair $(X,d)$ where
$X$ is a set, $d:X\times X \to \mathbb{R}$ is a real-valued
function, called the {\em distance function}, that satisfies,
$\forall\; \,x,y,z \in X$, properties
\begin{description}
  \item[(a)] $d(x,y)\geq 0$ and $d(x,y)=0$ iff $x=y$
  \item[(b)] $d(x,y)=d(y,x)$ (Symmetry property)
  \item[(c)] $d(x,z)\leq d(x,y)+d(y,z)$ (Triangle inequality).
\end{description}

Moreover, a {\em pseudo-metric space\/} is a pair $(X,\ell)$ where
$X$ is a set, $\ell:X\times X\to \mathbb{R}$ is a function, called
the {\em pseudo-distance function}, that satisfies, $\forall\;
\,x,y,z \in X$ (1') $\ell(x,y)\geq 0$ and $\ell(x,x)=0$ and
properties (2) and (3) of the distance function mentioned above.

Furthermore, there exists a canonical way of obtaining a metric
space from a pseudo-metric space. Consider a pseudo-metric space
$(X,\ell)$. Define on $X$ an equivalence relation, denoted by
$\sim$ here, s.t. $x\sim y$ iff $\ell(x,y)=0$. Now, let us denote
by $Y$ the set of all equivalence classes of $X$ under the
equivalence relation $\sim$. Further, define for $A, B\in Y$ a
function $e(A,B)=\ell(x,y)$ where $x\in A$ and $y\in B$. Then, the
function $e:Y\times Y \to \mathbb{R}$ is a distance function on
$Y$. Further, let $\Pi: X\to Y$ be the natural projection, {\em
ie}, for $x\in X$, $\Pi(x)=\{ y\in X| x\sim y\}$ an equivalence
class containing $x$. This projection function $\Pi$ is an
isometry, that is to say, it preserves the distance function $e$.

Then, it is now important to note that the {\em base space
$\mathbb{B}$ is a pseudo-metric space (and not a metric space)\/}
admitting a pseudo-metric (\ref{3d-metric-gen}), to be called as
the {\em Einstein pseudo-metric}, with the functions $P$, $Q$, $R$
being nowhere vanishing functions of their respective arguments,
and $x$, $y$, $z$ providing the ``coordinatization'' of
$\mathbb{B}$ s.t. $\mathbb{B}$ is a 3-dimensional
pseudo-Riemannian manifold.

Moreover, the restriction of the Einstein pseudo-metric
(\ref{3d-metric-gen}) on a P-set is a distance function. {\em
Therefore, we may, equivalently, define a P-set as that region of
the base space $\mathbb{B}$ over which the restriction of the
Einstein pseudo-metric (\ref{3d-metric-gen}) is a metric
function}. It is now obvious as to why a singleton set of
$\mathbb{B}$ cannot be a P-set.

As an aside, we note also that any averaging procedure is now
well-defined over a collection of P-sets. Therefore, we may, in a
mathematically meaningful way, talk about the ``concept of
energy-momentum tensor'' and some relation between the averaged
quantities, an ``equation of state'' defining appropriately the
``state of a physical system'' under consideration. Clearly,
different such physical concepts remain ``mathematically
meaningful'' in the present situation \footnote{This is, in a
definite sense, the field-theoretic comprehension of the
energy-momentum tensor.}. But, we shall not consider this issue in
the present paper.

\subsection{Physical space is Standard Borel}
Let us denote by $d$ the distance function obtainable in a
canonical manner from the Einstein pseudo-metric
(\ref{3d-metric-gen}).

Let us now consider the Einstein metric space $(\mathbb{B}, d)$. A
set $B_d(x_o, r)=\{x\in\mathbb{B} \,|\, d(x,x_o)< r\}$ is called
as an {\em open $r$-ball around fixed $x_o\in\mathbb{B}$}, where
$r$ is a positive real number.

We call a subset $A \subset B$ as {\em bounded\/} if $d/A\times A$
is a bounded function. We call the real number $\mathrm{sup} \{
d(x,y)\,|\, x,y \in A\}\,=\, \delta(A)$ as a {\em diameter of a
non-empty bounded set $A$}. A subset $A \subset \mathbb{B}$ is
called an {\em open subset\/} of $\mathbb{B}$ if $\forall \;
x_o\in A$, there exists a positive real number $r$ s.t.
$B_d(x_o,r)\subset A$. The collection $\Gamma$ of all open subsets
of $\mathbb{B}$ then provides the {\em metric topology\/} on
$\mathbb{B}$.

It is now easy to see that every P-set is a member of the metric
topology $\Gamma$ since a suitable open $r$-ball around each of
its points is contained in it and, hence, every P-set is open in
$(\mathbb{B}, \Gamma)$. However, note that every open set of
$(\mathbb{B}, \Gamma)$ is {\em not\/} a P-set of $(\mathbb{B},d)$
and that ``being a P-set of the Einstein space $\mathbb{B}$'' is
{\em not\/} a {\em topological property}.

Now, it is also easy to see that the space $\mathbb{B}$ is a
separable and compact metric space. But, every compact metric
space is complete \cite{kdjoshi}. Hence, the base space
$\mathbb{B}$ is a complete separable metric space. Note further
that $\mathbb{B}$ is uncountable.

Recall that a Topological Space $(X, \Upsilon)$ is called {\em
Polish\/} if there exists a metric $d$ on a non-empty set $X$ s.t.
the metric space $(X,d)$ is complete and separable, and the
topology $\Upsilon$, called a {\em Polish Topology}, is induced by
the metric $d$.

Now, by a {\em $\sigma$-algebra\/} $\mathcal{B}$ we mean a
non-empty collection $\mathcal{B}$ of subsets of $X$ which is
closed under countable unions and complements.

The intersection of any family of $\sigma$-algebras on $X$ is
again a $\sigma$-algebra. If $\mathcal{E}$ is any collection of
the subsets of $X$, then the intersection of all the
$\sigma$-algebras containing $\mathcal{E}$ is the smallest
$\sigma$-algebra containing $\mathcal{E}$ and is called the {\em
$\sigma$-algebra generated by the collection $\mathcal{E}$}.

A {\em Borel Structure (Borel space)\/} is a pair $(X, {\cal B})$
where $X$ is a non-empty set and ${\cal B}$ is a $\sigma$-algebra
of the subsets of $X$. If $A$ is a non-empty subset of $X$, then
the collection of all subsets of $X$ of the form $A\bigcap B,\;B
\in {\cal B}$ is a $\sigma$-algebra on $A$, called as the {\em
induced $\sigma$-algebra on $A$}, which we denote by $A\bigcap
{\cal B}$ or by ${\cal B}|_A$ or by ${\cal B}/A$.

Two Borel spaces $(X_1, {\cal B}_1)$ and $(X_2, {\cal B}_2)$ are
(Borel) isomorphic to each other if $\exists$ a one-one map $\phi$
of $X_1$ onto $X_2$, a {\em Borel isomorphism\/} of two Borel
spaces, s.t. $\phi\left( {\cal B}_1\right)={\cal B}_2$.

A Borel space Borel-isomorphic to the Borel space of an
uncountable complete separable metric space is called a {\em
Standard Borel Space\/} (SBS). A SBS is Borel-isomorphic to the
Borel space of the unit interval, $[0,1]$, equipped with the
$\sigma$-algebra generated by its usual topology.

A SBS, $(X, {\cal B})$, is therefore a Polish Topological Space
with the $\sigma$-algebra generated by open sets in $X$. We call
this $\sigma$-algebra the Borel $\sigma$-algebra of $X$ and we
often denote it as ${\cal B}_X$. Members of ${\cal B}_X$ are the
{\em Borel sets\/} of $X$.

Now, as noted before, the Einstein space $(\mathbb{B}, d)$ is an
uncountable, complete, separable metric space. Therefore, we are
dealing here with a {\em Polish Topological Space\/} \cite{trim6,
krp, weiss0} $(\mathbb{B}, \Gamma)$ where $\Gamma$ is the topology
defined earlier.

Then, if we construct the smallest $\sigma$-algebra of the subsets
of $\mathbb{B}$ that contains every open subset of $\mathbb{B}$,
that is to say, the Borel Set Structure on $\mathbb{B}$, then we
will be dealing with a SBS - the {\em Standard Borel Space\/} -
$(\mathbb{B},\,\mathcal{B})$. The Einstein-space or the physical
space $\mathbb{B}$ is therefore a SBS.

\section{Some relevant known results}
\subsection*{Set theoretic considerations}
In this connection, we firstly note the following results from the
descriptive set theory \cite{krp}:
\begin{itemize} \item Any set in ${\mathcal B}_{\mathbb{B}}$ is either countable
or has the cardinality $\mathbf{c}$ of the continuum. \item If $A,
B \in {\mathcal B}_{\mathbb{B}}$ are of the {\em same
cardinality}, then $A$ and $B$ are Borel isomorphic. \item If $Y$
is another complete separable metric space of the {\em same
cardinality\/} as $\mathbb{B}$ and ${\mathcal B}_Y$ is its Borel
$\sigma$-algebra, then $\left(\mathbb{B}, {\mathcal
B}_{\mathbb{B}}\right)$ and $(Y, {\cal B}_Y)$ are Borel
isomorphic.
\item From the above, it follows that if $A\in \mathcal{B}_{\mathbb{B}}$ and
$C\in \mathcal{B}_Y$ have the {\em same cardinality}, then the
Borel spaces $(A, A\bigcap \mathcal{B}_{\mathbb{B}})$ and $(C,
C\bigcap{\mathcal B}_Y)$ are Borel isomorphic.
\end{itemize}

Note also that the restriction of ${\cal B}$ in $(\mathbb{B},
{\cal B})$ to any of the Borel sets $A \in {\cal B}$ yields again
a new Borel space $(A, {\cal B}/A)$.

Then, two P-sets of the {\em same cardinality}, belonging either
to the same Einstein-space $(\mathbb{B}, d)$ or to two different
Einstein-spaces $(\mathbb{B}, d_1)$ and $(\mathbb{B}, d_2)$, are
Borel-isomorphic. Further, the restriction of ${\cal
B}_{\mathbb{B}}$ to a P-set yields a Borel space.

Note that a P-set is to be a physical particle and the physical
attributes of a particle are to be measures on a P-set. Surely,
any copies of a physical particle must be indistinguishable from
each other except for their spatial locations.

Then, the Borel-isomorphic equivalence of P-sets of the {\em same
cardinality\/} is consistent with relevant such physical
conceptions since the same values of measures exist on equivalent
such P-sets. This is then an indication of the internal
consistency of the present physical framework.

Now, let $\mathcal{N}$ be a collection of subsets of space
$\mathbb{B}$, $\mathcal{N}\subseteq {\cal B}$, s.t.
\begin{enumerate} \item $\mathcal{N}$ is closed under countable union
\item $A\in {\cal B}$ and $N\in \mathcal{N}$ implies that $A\bigcap N
\in \mathcal{N}$, \item for $N\in\mathcal{N}, \;
N^c=\mathbb{B}-N\notin \mathcal{N}$.\end{enumerate} We call
$\mathcal{N}$ the {\em $\sigma$-ideal}. Note that $\mathcal{N}$
may contain either $\emptyset$ or $\mathbb{B}$, but not the both.

If $\mathcal{E}$ is any collection of subsets of $\mathbb{B}$,
then there exists a smallest $\sigma$-ideal containing
$\mathcal{E}$, the intersection of all $\sigma$-ideals containing
$\mathcal{E}$. It is called the {\em $\sigma$-ideal generated by
$\mathcal{E}$\/} and is obtained by taking all sets of the form
$B\bigcap E$ with $B\in {\cal B}$, $E\in \mathcal{E}$ and taking
countable unions of such sets.

A subset $E$ of $\mathbb{B}$ is said to have the {\em property of
Baire\/} if $E$ can be expressed as a symmetric union of an open
set $G$ and a set $M$ of the first Baire category, {\em ie},
expressible as the union $E\equiv G\triangle M=(G-M)\bigcup(M-G)$.
If $E$ has the property of Baire, then so does its complement in
$\mathbb{B}$. A set of the first Baire category is a countable
union of nowhere dense sets.

Since $\mathbb{B}$ is a SBS, every Borel subset of $\mathbb{B}$
has the property of Baire since the $\sigma$-algebra of sets with
the property of Baire includes the Borel $\sigma$-algebra of
$\mathbb{B}$. The collection of subsets of $\mathbb{B}$ with the
property of Baire is a $\sigma$-algebra generated by open subsets
together with the subsets of the first Baire category. Subsets of
first Baire category in $\mathbb{B}$ form a $\sigma$-ideal in the
$\sigma$-algebra of sets with the property of Baire.

In particular, we note that every P-set is a Borel subset of the
Einstein space $(\mathbb{B}, \mathcal{B}, d)$ and, hence, has the
property of Baire. (When we want to emphasize that the base space
has metric $d$, we will use the notation $(\mathbb{B},
\mathcal{B}, d)$.)

For $A, B \in {\cal B}$, we write $A=B\;({\rm mod}\,\mathcal{N})$
if $A-B$ and $B-A$, both, belong to $\mathcal{N}$.

A subset $B\subset \mathbb{B}$, $B\in{\cal B}$, is said to be {\em
decomposable\/} if it is expressible as a union of two disjoint
sets from ${\cal B}-\mathcal{N}$. Clearly, every such decomposable
set belongs to ${\cal B}-\mathcal{N}$.

We say that the Borel $\sigma$-algebra ${\cal B}$ of subsets of
$\mathbb{B}$ satisfies the {\em countability condition\/} if every
collection of pairwise disjoint sets from ${\cal B}-\mathcal{N}$
is either finite or countably infinite.

Now, let $X$ be a set, $\mathfrak{P}(X)$ be its {\em power set\/}
(the set of all subsets of $X$) and $\mathcal{A}$ an algebra of
subsets of $X$. A {\em set function\/}
$m:\mathfrak{P}(X)\to\mathbb{R}$, defined on $\mathcal{A}$, is a
{\em finitely additive measure\/} if
\begin{description}
\item{(1)} $0 \leq m(A) \leq \infty\;\forall\;A \in \mathcal{A}$
\item{(2)} $m(\emptyset)=0$
\item{(3)} $m(A\bigcup B)=m(A)+m(B)$ if $A,B \in \mathcal{A}$ and
$A\bigcap B = \emptyset$. \end{description}

A finitely additive measure $m$ with an additional property that
\begin{description} \item{(4)} $m\left( \bigcup_{i=1}^{\infty} A_i\right) =
\sum_{i=1}^{\infty}m\left( A_i \right)$ for every pairwise
disjoint sequence $\{ A_i: i\in\mathbb{N}\}$ and
$\bigcup_{i=1}^{\infty}A_i\;\in\; \mathcal{A}$ \end{description}
is called a {\em Countably Additive Measure\/} or simply {\em
Measure}. Further, if all subsets of sets of measure zero are
measurable, a measure is said to be a {\em complete measure}.

If $m$ is a measure on $(X,\mathcal{A})$, then a set
$E\in\mathcal{A}$ is of {\em finite $m$-measure\/} if $m(E)<
\infty$; is of {\em $\sigma$-finite $m$-measure\/} if $\exists \;
\{E_i\},\;i\in\mathbb{N},\; E_i\in\mathcal{A}$ s.t.\ $E\subseteq
\bigcup_{i=1}^{\infty}E_i$ and $m(E_i)<\infty,\;\forall\;i\in
\mathbb{N}$. If $m(A),\;A\in\mathcal{A}$ is finite
($\sigma$-finite) then the measure $m$ is {\em finite
($\sigma$-finite) measure on $\mathcal{A}$}. A measure is {\em
totally finite or totally $\sigma$-finite\/} if $m(X)$ is finite
or $\sigma$-finite.

If $\mathcal{A}$ is a $\sigma$-algebra of subsets of $X$ and $m$
is a measure on it, then $(X, \mathcal{A})$ is a {\em measurable
space\/} and the members of $\mathcal{A}$ are {\em
$\mathcal{A}$-measurable sets}. We call $(X, \mathcal{A}, m)$ as a
{\em Measure Space}.

We also define a {\em signed measure\/} as an extended,
real-valued, countably additive set function $\mu$ on the class,
$\mathcal{A}$, of all measurable sets of a measurable space
$(X,\mathcal{A})$ s.t.\ $\mu(\emptyset)=0$, and s.t.\ $\mu$
assumes at most one of the values $+\infty$ and $-\infty$.

If $\mu$ is a signed measure on a measurable space
$(X,\mathcal{A})$, we call a set $E$ {\em positive (negative) wrt
$\mu$\/} if, $\forall\;F\in\mathcal{A}$, $E\bigcap F$ is
measurable and $\mu(E\bigcap F)\geq 0$ ($\mu(E\bigcap F)\leq 0$).
The empty set is both $\mu$-positive and $\mu$-negative in this
sense.

If $\mu$ is a signed measure on $(X,\mathcal{A})$, then there
exist two disjoint sets $A,B\in \mathcal{A}$ s.t.\ $A\bigcup B=X$
and $A$ is $\mu$-positive while $B$ is $\mu$-negative. The sets
$A$ and $B$ are said to form the {\em Hahn Decomposition of $X$
wrt $\mu$}. It is not unique.

For every $E\in \mathcal{A}$, we define $\mu^+(E)=\mu(E\bigcap
A)$, the {\em upper variation of $\mu$}, and
$\mu^-(E)=\mu(E\bigcap B)$, the {\em lower variation of $\mu$},
and $|\mu|\,(E)=\mu^+(E)+\mu^-(E)$, the {\em total variation of
$\mu$}, where $A,B$ are as in the Hahn decomposition. [Note that
$|\mu(E)|$ and $|\mu|(E)$ are not the same.]

The upper, the lower and the total variations (of $\mu$) are
measures and $\mu(E)=\mu^+(E)-\mu^-(E)$
$\forall\;\,E\in\mathcal{A}$, the {\em Jordon decomposition}. If
$\mu$ is finite or $\sigma$-finite, then so are $\mu^+$ and
$\mu^-$; at least one of $\mu^+$ and $\mu^-$ is always finite.

A {\em simple function on $(X,\mathcal{A})$ \/} is
$f=\sum_{i=1}^n\alpha_i \chi_{_{E_i}}$ where $E_i\in \mathcal{A}$,
$\chi_{_{E_i}}$ is the {\em characteristic function of the set
$E_i$\/} and $\alpha_i\in\mathbb{R}$. This simple function $f$ is
{\em $\mu$-integrable\/} if $\mu(E_i)< \infty$ $\forall \;i$ for
which $\alpha_i\neq 0$. The {\em $\mu$-integral of $f$\/} is $\int
f(x)d\mu(x)\;\mathrm{or}\; \int
f\,d\mu=\sum_{i=1}^n\alpha_i\mu(E_i)$.

If, $\forall\;\epsilon>0$, $\lim_{n\to\infty}$ $m\left( \{x\in X:
|f_n(x)\right.$$-f(x)|\geq \epsilon$$\left.\}\right)= 0$, a
sequence $\{f_n\}$ of a.e.\ finite-valued measurable functions is
said to {\em converge in measure\/} to a measurable function $f$.

Given two integrable simple functions $f$ and $g$ on a measure
space $(X,\mathcal{A})$, define now a pseudo-metric
$\rho(f,g)=\int |f-g|\;d\mu$. A sequence $\{f_n\}$ of integrable
simple functions is {\em mean fundamental\/} if $\rho(f_n,g_m)\to
0$ if $n,m\to\infty$.

An a.e.\ finite-valued, measurable function $f$ on
$(X,\mathcal{A})$ is {\em $\mu$-integrable\/} if there is a mean
fundamental sequence $\{f_n\}$ of integrable simple functions
which converges in measure to $f$.

Lebesgue-Radon-Nikodym (LRN) theorem \cite{trim6, halmos1} states
that: If $(X,\mathcal{A},m)$ is a totally $\sigma$-finite measure
space and if a $\sigma$-finite measure $\nu$ on $\mathcal{A}$ is
absolutely continuous wrt $m$, then $\exists$ a finite valued
measurable function $f$ on $X$ s.t.\ $\nu(E)=\int_{_E}f\,d\mu$ for
every measurable set $E\in\mathcal{A}$. The function $f$ is unique
in the sense that if also $\nu(E)=\int_{_E}g\,d\mu$, then $f=g\;
(\mathrm{mod}\;\mu)$, {\it ie}, equality holding modulo a set of
$\mu$-measure zero or $\mu$-a.e.

If $\mu$ is a totally $\sigma$-finite measure and if
$\nu(E)=\int_{_E}f\,d\mu$ $\forall \;\;E\in\mathcal{A}$, we write
$f=\frac{d\nu}{d\mu}$ or $d\nu=f\,d\mu$. We call
$\frac{d\nu}{d\mu}$ the LRN-derivative and all the properties of
the differential formalism hold for it, importantly, $\mu$-a.e.

A set $E\subseteq X$ is $G_{\delta}$ if $E=\bigcap_{i=1}^{\infty}
U_i, \;U_i$ open in $X$; it is $F_{\sigma}$ if $E=
\bigcap_{i=1}^{\infty}C_i,\;C_i$ closed in $X$. Class of all
$G_{\delta}$ ($F_{\sigma}$) sets is closed under finite unions and
countable intersections.

For a locally compact Hausdorff space $X$, let $\mathcal{C}$ be
the class of all compact subsets of $X$, $\mathcal{A}$ be the
$\sigma$-algebra generated by $\mathcal{C}$, and $\mathcal{U}$ be
the class of all open sets belonging to $\mathcal{A}$. Also, let
$\mathcal{C}_o$ denote the class of all compact subsets of $X$
which are $G_{\delta}$, $\mathcal{A}_o$ be the $\sigma$-algebra
generated by $\mathcal{C}_o$, and $\mathcal{U}_o$ be the class of
all open sets belonging to $\mathcal{A}_o$.

Members of $\mathcal{A}_o$ are the {\em Baire sets of $X$} and
$\mathcal{U}_o$ is the class of all open Baire sets. A real-valued
function on $X$ is {\em Baire measurable\/} or simply a {\em Baire
function\/} if it is $\mathcal{A}_o$-measurable. A measure on
$\mathcal{A}_o$ is called a {\em Baire measure}. If
$\mathcal{A}=\mathcal{B}_{_X}$, then the members of $\mathcal{A}$
are {\em Borel measurable\/} and measure on $\mathcal{A}$ is a
{\em Borel measure}. A real-valued $\mathcal{B}_{_X}$-measurable
function is called a Borel-measurable function or simply a {\em
Borel function}.

A {\em Haar measure\/} is a Borel measure $\mu$ in a locally
compact topological group $X$ s.t.\ $\mu(U)>0$ for every non-empty
Borel open subset $U$ of $X$, and $\mu(x\,E)=\mu(E)$ for every
Borel set $E$ of $X$. A Haar measure is a translation-invariant
Borel measure which is not identically zero.

An {\em atom of a measure $m$\/} is an element $E\in \mathcal{A},
\; E\neq\emptyset$ s.t.\ if $F\subset E$, then either $m(F)=0$ or
$m(F)=m(E)$. A measure with no atoms is {\em non-atomic}. A
measure space with non-atomic measure on it is non-atomic. We also
call a finite measure $\nu$ on $\mathcal{A}$ {\em purely atomic\/}
if there exists a countable set $C\in\mathcal{A}$ s.t.\
$\nu(X-C)=0$.

A measure $m$ is called {\em regular\/} if given $A\in
\mathcal{A}$ and $\epsilon>0$, there exist $G,F\in \mathcal{A}$
s.t.\ $m(G)-\epsilon\leq m(A)\leq m(F)+\epsilon$.

For $(X,\mathcal{A},\mu)$ a measure space, let $\mathfrak{S}(\mu)$
be the set of all measurable sets with finite $\mu$-measure. For
any $E,F\in\mathfrak{S}(\mu)$, let $\rho(E,F)=\mu(E\triangle F)$.
The function $\rho$ so defined is a metric on $\mathfrak{S}(\mu)$
and the metric space $(\mathfrak{S}(\mu),\rho)$ is called the {\em
metric space of or associated to $(X,\mathcal{A},m)$}.

A SBS equipped with a finite or $\sigma$-finite measure $m$ is
called as the {\em Standard Measure Space\/}  (SMS). If, in
particular, $m(X)=1$, a SMS is called as the {\em Standard
Probability Space\/} (SPS) and $m$ is called as the {\em
probability measure on $\mathcal{B}$}.

Now, we note that a P-set is an uncountable separable complete
metric space with the restriction of the Einstein pseudo-metric
(\ref{3d-metric-gen}) to it as a metric. Therefore,
Borel-isomorphic P-sets of the {\em same cardinality\/} are
Borel-isomorphic SMSs/SPSs. Such P-sets,  as particles of the {\em
same species}, then possess measures as indistinguishable physical
attributes.

Indeed, it can be called a deep mystery of the micro-cosmos as to
how all the elementary particles of the same type possess exactly
identical physical attributes such as mass, charge etc. Perhaps,
the above characterization of physical particles as P-sets of the
Einstein space $(\mathbb{B}, d)$ is a solution to this
ill-understood deep mystery.

In the above, we dealt with the, descriptive or otherwise, set
theoretic properties of the Einstein space $\mathbb{B}$. In order
to discuss the dynamics of this space, and that of the P-sets, we
find the following mathematical formalism useful.

\subsection*{Dynamical considerations}
For an extended real-valued function $f:X\to \mathbb{R}$, the set
$\{x\in X\;|\; f(x)\neq 0\}=\mathfrak{Support}\,(f)$ is a {\em
support of $f$ on $X$}. A function $f:X\to \mathbb{R}$ on a
measurable space $X$ is a {\em measurable function\/} if
$\mathfrak{Support}\,(f)\bigcap f^{-\,1}(M)$ is a measurable set
where $M$ is any Borel subset of $\mathbb{R}$.

The forward image of a measurable subset of $X$ under a measurable
function $f$ {\em need not be\/} measurable, in general.

{\em Lusin's theorem}: If $f$ is a measurable function from a SBS
into another SBS and if $f$ is {\em countable to zero\/}, {\em
ie}, if the inverse image of every singleton set is at most
countable, then the forward image under $f$ of a Borel set is
Borel.

Now, a one-one measurable map $T$ of a Borel space $(X, {\cal B})$
onto itself s.t.\ $T^{-\,1}$ is also measurable is called a {\em
Borel automorphism}. That is to say, a Borel automorphism of $(X,
{\cal B})$ is a one-one and onto map $T:X\to X$ s.t. $T(B)\in
{\cal B}\;\forall\; B \in \mathcal{B}$. We call a set $A$ {\em
invariant under $T$\/} or {\em $T$-invariant\/} if $TA=A$.

An automorphism of $X$ onto $X$ is, in general, not a Borel
automorphism. But, if $(X, {\cal B})$ is a SBS then a measurable
one-one map of $X$ onto $X$ is a Borel automorphism.

The Einstein-space $\mathbb{B}$ is a SBS. Hence, we will be
considering only the Borel automorphisms of $(\mathbb{B},
\mathcal{B})$ and the class of Borel measures on it. This class
will be that of physical attributes of the particle associated
with a P-set.

Then, the Ramsay-Mackey theorem states that: If $T:X\to X$ is a
Borel automorphism of a SBS $(X,\mathcal{B})$, then there exists a
topology $\Gamma$ on $X$ s.t.
\begin{description} \item{(a)} $(X, \Gamma)$ is a complete,
separable, metric space \item{(b)} Borel sets of $(X, \Gamma)$ are
precisely those in $\mathcal{B}$ \item{(c)} $T$ is a homeomorphism
of $(X, \Gamma)$.\end{description} Note that $X$ is {\em same\/}
for $(X,\mathcal{B})$ and $(X,\Gamma)$. However, this is not the
usual formalism of the topological dynamics.

Alternatively, if $X$ is the underlying set and if $T$ is a Borel
automorphism on a SBS $(X, \mathcal{B})$ and $\mathcal{C}\subseteq
\mathcal{B}$ is a countable collection, then there exists a
complete separable metric topology, {\em ie}, Polish topology,
$\Gamma$, on $X$ such that \begin{description}
\item{(1)} $T$ generates the $\sigma$-algebra $\mathcal{B}$ \item{(2)} $T$ is
a homeomorphism of $(X, \Gamma)$ \item{(3)} $\mathcal{C} \subseteq
\Gamma$, and lastly, \item{(4)} $\Gamma$ has a clopen base, {\em
ie}, sets which are both open and closed in $\Gamma$.
\end{description}

If $\Gamma_o$ is a Polish topology on $X$ which generates
$\mathcal{B}$, $\{\Gamma_i,\;i\in \mathbb{N}\}$ are Polish
topologies on $X$ with $\Gamma_o \subseteq \Gamma_i \subseteq
\mathcal{B}$ then $\exists$ a Polish topology
$\Gamma_{\infty}\;\left( \subseteq \mathcal{B}\right)$ such that
$\bigcup_{i=1}^{\infty} \Gamma_i \subseteq \Gamma_{\infty}$ and
$\Gamma_{\infty}$ is the topology generated by all {\em finite
intersections\/} of the form $\bigcap_{i=1}^n\,G_i,\;G_i\in
\Gamma_i$ for $i, n \in \mathbb{N}$.

Further, given $B\in\mathcal{B}$, there exists a Polish topology
$\bar{\Gamma}$, $\Gamma_o\subseteq\bar{\Gamma}\subseteq
\mathcal{B}$ s.t.\ $B\in\bar{\Gamma}$. Moreover, $\bar{\Gamma}$
can be chosen to have a clopen base.

Now, as noted, the Einstein space $\mathbb{B}$ is a Polish space
with the topology generated by the Einstein pseudo-metric
(\ref{3d-metric-gen}) being the Polish topology on $\mathbb{B}$.
Also, every P-set is open in $\mathbb{B}$ but every open set of
$\mathbb{B}$ is {\em not\/} a P-set.

However, it is clear that given any open subset $A\subset
\mathbb{B}$, there is a Einstein pseudo-metric
(\ref{3d-metric-gen}) s.t.\ the set $A$ is a P-set, that is, the
restriction of the Einstein pseudo-metric to $A\subset \mathbb{B}$
is a distance function. To achieve this, we select appropriate
functions $P$, $Q$, $R$ in (\ref{3d-metric-gen}) to make the
subset $A$ a P-set. Then, given $B\in \mathcal{B}$, there exists a
suitable Einstein pseudo-metric (\ref{3d-metric-gen}) generating
the topology $\bar{\Gamma}$ for which $B$ is a open set.

That the Einstein pseudo-metric (\ref{3d-metric-gen}) can be
defined on a SBS and the same pseudo-metric is a distance function
on certain Borel subsets of a SBS, the P-sets, is a result of this
paper.

We also note that the restriction of a Polish topology to a
$G_{\delta}$ set is also a Polish topology. Furthermore, for any
countable collection $(B_j)_{j=1}^{\infty}\subseteq \mathcal{B}$,
there exists a Polish topology $\Gamma$ (which can be chosen to
have a clopen base) s.t.\ $\Gamma_o\subseteq \Gamma \subseteq
\mathcal{B}$ and for all $j$, $B_j\in\Gamma$.

An extension of this result for {\em jointly measurable flows\/}
is available in \cite{vmw}. A group $T_t,\;t\in \mathbb{R}$ of
Borel automorphisms on a SBS is a {\em jointly measurable flow\/}
if \begin{description}
\item{(1)} the map $(T, x) \mapsto T_t x$ from $\mathbb{R}\times X \to
X$ is measurable, where $\mathbb{R}\times X$ is endowed with the
usual product Borel structure \item{(2)} $T_0x=x \;\forall\;x \in
X$ and, \item{(3)} $T_{t+s}x=T_t \circ T_sx$ for all $t,s \in
\mathbb{R}$ and for all $x\in X$. \end{description}

Further, if $T$ is a homeomorphism of a Polish space $X$ then
there exists a compact metric space $Y$ and a homeomorphism $\tau$
of $Y$ s.t.\ $T$ is isomorphic as a homeomorphism to the
restriction of $\tau$ to a $\tau$-invariant $G_{\delta}$ subset of
$Y$. We can choose the $\tau$-invariant set to be dense in $Y$.
This result is due to N. Krylov and N. Bogoliouboff.

Combined with the theorem of Ramsay and Mackey, this shows that a
Borel automorphism on a SBS can be viewed as a restriction of a
homeomorphism of a compact metric space to an invariant
$G_{\delta}$ subset.

Furthermore, in \cite{varadrajan}, it is proved that: Given a
jointly measurable action of a second countable locally compact
group $G$ on a SBS $(X,\mathcal{B})$ there is a compact metric
space $Y$ on which $G$ acts continuously and a Borel subset
$X'\subseteq Y$ which is $G$-invariant and s.t.\ the $G$-actions
on $X'$ and $X$ are isomorphic.

Moreover, Becker \cite{trim6} has proved: If $G$ is a Polish group
acting continuously on a Polish space $(X,\Gamma)$ and if
$\mathcal{C}$ is a countable family of $G$-invariant Borel subsets
of $X$, then there exists a Polish topology $\Gamma\supseteq
\Gamma$ s.t.\ \begin{description}\item{(1)} $G$ acts continuously
on $(X,\Gamma_1)$ \item{(2)} $\Gamma_1$ generates the same Borel
structure $\mathcal{B}$ on $X$ as $\Gamma$ \item{(3)} Every set in
$\mathcal{C}$ is $\Gamma_1$ closed.\end{description}

The generalization \cite{trim6} of the Ramsay-Mackey theorem to
locally compact second countable group actions (due to Kechris)
states that: Let a second countable locally compact group $G$ act
in a jointly measurable fashion on a SBS $(X,\mathcal{B})$. Then,
we can imbed $X$ as a $G$-invariant Borel set in a compact metric
space $Y$. We can then enlarge the topology of $Y$ to be a Polish
topology $\Gamma_1$ wrt which the $G$-action remains jointly
continuous and s.t.\ $X$ is closed under $\Gamma_1$. If we now
restrict $\Gamma_1$ to $X$ then $X$ is Polish under this topology
and the $G$-action on $X$ is jointly continuous.

The Glimm-Effros Theorem \cite{trim6} states that: If $X$ is a
complete separable metric space and $G$ a group of homeomorphisms
of $X$ onto itself s.t.\ for some non-isolated point $x\in X$, the
set $Gx$, the orbit of $x$ under $G$, is dense in $X$, then there
is a continuous probability measure $\mu$ on Borel subsets of $X$
s.t.\ every $G$-invariant Borel set has measure zero or one.

Now, we say that a group $G$ of homeomorphisms of a Polish space
$X$ admits a {\em recurrent point\/} $x$ if there exists a
sequence $(g_n)_{n=1}^{\infty}$ of elements in $G$ s.t.\ $g_nx\neq
x\;\forall\;n$ and $g_nx\to x$ as $n\to\infty$. A recurrent point
$x$ is not isolated in the closure of $Gx$ and its $G$-orbit is
clearly dense in the closure of $Gx$. It is easy to see that if
the quotient topology on $X/G$ is not $T_o$ then the $G$-action on
$X$ admits a recurrent point.

A Borel set $W$ is said to be {\em $G$-wandering\/} if the sets
$gW,\;g\in G$ are pairwise disjoint. We write $\mathcal{W}_{_G}$
for the $\sigma$-ideal generated by $G$-wandering Borel sets, it
consists of {\em countable unions\/} of $G$-wandering sets of $X$.

Then, in this case, we have the result that: If a group of
homeomorphisms $G$ of a Polish space $X$ acts {\em freely\/} and
does not admit a recurrent point, then $X\in \mathcal{W}_{_G}$.

That a P-set is {\bf\it not\/} a singleton set of $\mathbb{B}$ is,
of course, consistent with all such results.

Precisely, some of the above results guarantee the existence of
P-sets on the base space $\mathbb{B}$. Also, certain theorems
\cite{trim6} establish the existence of a continuous finite
measure on $\mathcal{B}$.

Now, since a SBS is Borel-isomorphic to the Borel space of the
unit interval $X=[0,1]$ equipped with the $\sigma$-algebra
generated by its usual topology, we can restrict our discussion to
it.

Then, for any $x\in X$, we define the {\em orbit of $x$ under
$T$\/} as the set $\{T^nx|n\in \mathbb{Z}\}$. We call a point
$x\in X$ a {\em periodic point of $X$\/} if $T^nx=x$ for some
integer $n$ and call the smallest such integer the {\em period of
$x$ under $T$}.

For $A\subseteq X$ and $x\in A$, we say that the point $x$ is {\em
recurrent in $A$\/} if $T^nx\in A$ for infinitely many positive
(fimp) $n$ and for infinitely many negative (fimn) $n$ and we call
the point $x$ a {\em recurrent point}. For a metric space $(X,d)$,
a point $x\in X$ is recurrent if $\liminf_{n\to\infty}
d(x,T^nx)=0$.

Two Borel automorphisms, $T_1$ on a Borel space $(X_1,
\mathcal{B}_1)$ and $T_2$ on a Borel space $(X_2, \mathcal{B}_2)$,
are said to be {\em isomorphic\/} if there exists a Borel
isomorphism $\phi: X_1\to X_2$ s.t.\ $\phi\,T_1\,\phi^{-\,1}=T_2$.

We also say that Borel automorphisms $T_1$ and $T_2$ are {\em
weakly equivalent\/} or {\em orbit equivalent\/} if there exists a
Borel automorphism $\phi: X_1\to X_2$ s.t.\ $\phi\left(
\mathrm{orb}(x,T_1)\right)=\mathrm{orb}\left( \phi(x),
T_2\right)$, $\forall$ $\;x\in X_1$.

If two Borel automorphisms are isomorphic then they are also
orbit-equivalent. However, the converse is, in general, not true.

Now, we say that a Borel automorphism $T$ is an {\em elementary
Borel automorphism\/} or that the {\em orbit space of $T$ admits a
Borel cross-section\/} or that {\em $T$ admits a Borel
cross-section\/} iff there exists a measurable set $B$ which
intersects each orbit under $T$ in exactly one point.

Clearly, if $n$ is the period of $x$ under $T$, then the set $\{
x, Tx, T^2x, ..., T^{n-1}x\}$ consists of {\em distinct\/} points
of $X$. Now, for every positive integer $n$, let $E_n=\{ x\, |\,
Tx\neq x, ..., T^{n-1}x\neq x, T^nx=x\}$, and $E_{\infty}=\{x\,|\,
T^nx\neq x \;\mathrm{for \;all\;integers}\;n \}$. Then, each
$E_n,\;n<\infty,$ is Borel, $E_m\bigcap E_n=\emptyset$ if $m\neq
n$, and $\bigcup_{n=1}^{\infty}E_n=X$. Clearly, each $E_n$ is a
$T$-invariant Borel subset of $X$.

Now, if $y\in \{x, Tx, ..., T^{n-1}x\}$, then we clearly see that
$\{ x, Tx, ..., T^{n-1}x\}=\{y, Ty, ..., T^{n-1}y\}$. Moreover,
due to the natural order on $[0,1]$, if $y=\mathrm{min} \{ x, Tx,
..., T^{n-1}x\}$, then $y<Ty$, $y<T^2y$, ..., $y<T^{n-1}y$,
$y=T^ny$. Then, we can define $B_n=\{ y\in E_n \,|\, y<Ty,\,
...,\, y<T^{n-1}y\}$.

Then, for $n<\infty$, $B_n$ is a measurable subset of $E_n$ and it
contains exactly one point of the orbit of each $x\in E_n$. Note,
however, that $B_{\infty}$ need not be measurable.

Now, $X-E_{\infty}=\bigcup_{n=1}^{\infty}\bigcup_{k=0}^{n-1}
T^kB_n$. The set $B=\bigcup_{k=1}^{\infty} B_k$ is Borel and has
the property that orbit of any point in $X-E_{\infty}$ intersects
$B$ in exactly one point. Let $\mathbf{c}_n(T)$ denote the
cardinality of $B_n$, $n<\infty$. The sequence of integers
$\{\mathbf{c}_{\infty}(T),\, \mathbf{c}_1(T),\,\mathbf{c}_2(T),\,
...\}$ is called the {\em cardinality sequence associated to $T$}.

If $T_1$ and $T_2$ are orbit equivalent, then their associated
cardinality sequences are the same. Also, if $T_1$ and $T_2$ are
elementary and the associated cardinality sequences are the same,
then $T_1$ and $T_2$ are isomorphic and orbit equivalent.

A measurable subset $W\subset X$ is {\em $T$-wandering\/} or {\em
wandering under $T$\/} if $T^nW,\;n\in\mathbb{Z}$, are pairwise
disjoint. Clearly, a wandering set intersects the orbit of any
point in at most one point, it never intersects the orbit of a
periodic point.

The $\sigma$-ideal generated by all $T$-wandering sets in
$\mathcal{B}$ will be denoted by $\mathcal{W}_{_T}$ and will be
called a Shelah-Weiss ideal of $T$ \cite{weiss0}.

Note that if $T$ is a homeomorphism of a separable metric space
$(X,d)$ and $T$ has no recurrent points then $\mathcal{W}_{_T} =
\mathcal{B}$, {\it ie}, there is a wandering set $W$ s.t.\
$X=\bigcup_{n=-\infty}^{\infty}T^nW$.

A subset $A\subset \mathrm{orb}(x,T)$ is called {\em bounded
below} ({\em bounded above}) if the set of integers $n$ s.t.\
$T^nx\in A$ is bounded below (bounded above). A subset $A\subset
\mathrm{orb}(x,T)$ is called {\em bounded\/} iff it is both
bounded above and below. A set which is not bounded is called {\em
unbounded}.

A sufficient condition for a set $N\in \mathcal{B}$ to be a
$T$-wandering set, {\em ie}, a sufficient condition for
$N\in\mathbb{B}$ to belong to $\mathcal{W}_{_T}$, is that
$\forall\; x\in X$, $N\bigcap \mathrm{orb}(x,T)$ is either bounded
above or below.

One of the very basic results of the study of Borel automorphisms
is:

{\em Poincar\'{e} Recurrence Lemma}: Let $T$ be a Borel
automorphism of a SBS $(X,\mathcal{B})$. Then, given
$A\in\mathcal{B}$ $\exists\;N\in \mathcal{W}_{_T}$ s.t.\
$\forall\;x\in A_o=A-N$ the points $T^nx$ return to $A$ fimp $n$
and fimn $n$.

Now, note also that if $x\in A_o=A-N$ then $T^kx$ returns to $A_o$
fimp $k$ and fimn $k$ because $N$ is $T$-invariant and $x\notin
N$.

Also, if $A\in \mathcal{B}$, and if $A_o=A-N$ is as in the
Poincar\'{e} Recurrence Lemma, then $\bigcup_{k=-\infty}^{\infty}
T^kA=\bigcup_{k=0}^{\infty}T^kA \; (\mathrm{mod}\;
\mathcal{W}_{_T})$.

Now, suppose that $\mathcal{N}\subseteq \mathcal{B}$ is a
$\sigma$-ideal s.t.\ $T\mathcal{N}=T^{-\,1}\mathcal{N} =
\mathcal{N}$ and $\mathcal{W}_{_T}\subseteq \mathcal{N}$. Clearly,
given $A \in \mathcal{B}$, $\exists\;N\in \mathcal{N}$ s.t.\
$\forall\; x\in A_o=A-N$, $T^nx$ returns to $A_o$ fimp $n$ and
fimn $n$.

Of particular interest to us is a finite or $\sigma$-finite
measure $m$ on $\mathcal{B}$. The $\sigma$-ideal of $m$-null sets
in $\mathcal{B}$ will be denoted by $\mathcal{N}_m$.

A Borel automorphism $T$ is said to be {\em dissipative wrt $m$\/}
if there exists a $T$-wandering set $W$ in $\mathcal{B}$ s.t.\ $m$
is supported on $\bigcup_{n=-\infty}^{\infty}T^nW$.

On the other hand, a Borel automorphism $T$ is {\em conservative
wrt $m$\/} or {\em $m$-conservative\/} if $m(W)=0$ $\forall$
$T$-wandering sets $W\in\mathcal{B}$. Clearly, for any
$m$-conservative $T$, $\mathcal{W}_{_T}\subseteq \mathcal{N}_m$.

Poincar\'{e} Recurrence Lemma for $m$-conservative $T$: If $T$ is
$m$-conservative and if $A\in \mathcal{B}$ is given, then for
almost every (f.a.e.) $x\in A$ the points $T^nx$ return to $A$
fimp $n$ and fimn $n$.

Further, if $m$ is a probability measure on $\mathcal{B}$, {\em
ie}, $m(X)=1$, and is $T$-invariant, {\em ie}, $m\circ
T^{-\,1}=m$, then $\mathcal{W}_{_T}\subseteq \mathcal{N}_m$,
$T\mathcal{N}_m =T^{-\,1}\mathcal{N}_m = \mathcal{N}_m$.

Poincar\'{e} Recurrence Lemma (measure theoretic): If a Borel
automorphism $T$ on $(X,\mathcal{B})$ preserves a probability
measure on $\mathcal{B}$, and if $A\in \mathcal{B}$ is given, then
f.a.e. $x\in A$ the points $T^nx$ return to $A$ fimp $n$ and fimn
$n$.

Poincar\'{e} Recurrence Lemma (Baire Category): If $T$ is a
homeomorphism of a complete separable metric space $X$ which has
no $T$-wandering non-empty open set, then for every $A\subseteq X$
with the property of Baire (in particular, for any Borel set $A$)
there exists a set $N$ of the first Baire category (which is Borel
if $A$ is Borel) such that for each $x\in A-N$, the points $T^nx$
return to $A-N$ fimp $n$ and fimn $n$.

Now, a measure-preserving automorphism $T$ on a SPS $(X,
\mathcal{B}, \mu)$ is a {\em Bernoulli-Shift\/} or {\em B-shift\/}
if there exists a finite or a countably infinite partition
$\mathcal{P}=\{P_1,\;P_2\;...\}$ of $X$ into measurable sets s.t.\
\begin{description} \item{(a)}
$\bigcup_{n=-\infty}^{\infty}T^n\mathcal{P}$ generates
$\mathcal{B}_X$ up to $\mu$-null sets \item{(b)} the family
$\{T^n\mathcal{P}\,|\, n\in \mathcal{Z}\}$ is independent in the
sense that for all $k$, for all distinct integers $n_1$, $n_2$,
..., $n_k$, and for all $P_1$, $P_2$, ..., $P_{i_{k}}$ $\in
\mathcal{P}$, the sets $T^{n_1}P_{i_{1}}$, $T^{n_2}P_{i_{2}}$,
..., $T^{n_k}P_{i_{k}}$ are independent, {\em ie}, $\mu\left(
T^{n_1}P_{i_{1}}\bigcap ... \bigcap T^{n_k}P_{i_{k}} \right) =
\prod_{j=1}^k \mu\left(T^{n_j}P_{i_{j}}\right)$ which in view of
the measure preserving character of $T$ is equal to
$\mu(P_{i_1})...\mu(P_{i_k})$.\end{description} We call the
partition $\mathcal{P}$ satisfying the above an {\em independent
generator of $T$}.

$T$ is {\em $m$-deterministic\/} (otherwise, {\em non
deterministic\/}) if $\forall \; n$,
$\mathcal{P}_n=\mathcal{P}_{n+1}\;(\mathrm{mod}\;m)$ in that,
given $A\in \mathcal{P}_n$, $\exists\;B\in \mathcal{P}_{n+1}$
s.t.\ $m\left( A\triangle B\right)=0$. If $T$ is deterministic,
$\mathcal{P}_n = \mathcal{P}_k\,(\mathrm{mod}\,m),\;\forall\;n,k$.

A non-deterministic Borel automorphism $T$ is a {\em Kolmogorov
Shift\/} or {\em K-shift\/} if
$\bigcap_{n=-\infty}^{\infty}\mathcal{P}_n$ consists of sets with
probability zero or one.

A B-shift is a K-shift and, hence, is of non-deterministic nature
in the above sense.

Now, a measure preserving Borel automorphism $T$ on a probability
space $(X, \mathcal{B}, m)$ is said to be {\em ergodic\/} if for
every $T$-invariant $A\in \mathcal{B}$, $m(A)=0$ or $m(X-A)=0$.
Note that such a $T$ is ergodic iff every real-valued measurable
$T$-invariant function $f$ is constant a.e.

Now, if $T$ is measure-preserving, ergodic and for some singleton
$\{x\}\in \mathcal{B}$, $m\left( \{x\}\right)>0$, then $x$ must be
a periodic point of $T$. A non-trivial measure-preserving ergodic
system is therefore the one for which $m$ is non-atomic.

The system $\left( X, {\cal B}, \mathcal{N}, T \right)$ is called
as a {\em descriptive dynamical system\/} \cite{trim6}.

Now, $T$ is said to be {\em descriptively ergodic\/} or that $T$
is said to {\em act in a descriptively ergodic manner\/} if
$T\mathcal{N}=\mathcal{N}$ and if $TA=A,\;A\in{\cal B}$ implies
either $A\in \mathcal{N}$ or $X-A\in \mathcal{N}$.

Nadkarni's theorem \cite{trim6} states: if $(X, {\cal B},
\mathcal{N}, T)$ is a descriptive dynamical system s.t.
\begin{description} \item {(a)} every member of ${\cal B}-\mathcal{N}$
is decomposable \item{(b)} ${\cal B}$ satisfies the countability
condition \item{(c)} $T$ is descriptively ergodic \item{(d)} $X$
is bounded,\end{description} then there exists a finite measure
$\mu$ on ${\cal B}$ s.t.
\begin{description}\item{(1)} $\mathcal{N}\;=\;\left\{ B\in{\cal B}: \mu(B)=0
\right\}$ \item{(2)} $\mu$ is continuous \item{(3)} $T$ is
$\mu$-measure preserving, and \item{(4)} $T$ is ergodic, {\em ie},
$TA=A,\; A\in {\cal B}$ implies that $\mu(A)=0$ or $\mu(X-A)=0$.
\end{description}

As a corollary of this theorem, we also have: Let $(X,\mathcal{B},
\mathcal{N},T)$ be a descriptive dynamical system s.t.\
\begin{description} \item{(a)} every member of $\mathcal{B}-
\mathcal{N}$ is decomposable, \item{(b)} $\mathcal{B}$ satisfies
the countability condition \item{(c)} $T$ is descriptively
ergodic, \item{(d)} $\exists\;B\in\mathcal{B}- \mathcal{N}$ which
is bounded \end{description} Then, there exists a unique
continuous $\sigma$-finite measure $m$ on $\mathcal{B}$ s.t.\ its
null sets in $\mathcal{B}$ form precisely the ideal $\mathcal{N}$
and $T$ is ergodic and measure preserving wrt $m$.

Furthermore, it can also be shown \cite{trim6} that: for a SBS
$(X,\mathcal{B})$ and $T:X\to X$ a Borel automorphism, there
exists a finite continuous measure $m$ on $\mathcal{B}$ so that
$T$ is {\em non-singular\/} and ergodic iff there exists a
$\sigma$-ideal $\mathcal{N}\subseteq \mathcal{B}$ s.t.\ the system
$(X,\mathcal{B},\mathcal{N},T)$ has the following properties
\begin{description} \item{(i)} every member of $\mathcal{B}-
\mathcal{N}$ is decomposable \item{(ii)} $\mathcal{B}$ satisfies
the countability condition \item{(iii)} $T$ is descriptively
ergodic \item{(iv)} $\exists\;B\in \mathcal{B}- \mathcal{N}$ which
is bounded \end{description}

Now, let $X=\{0,1\}^{\mathbb{N}}$, the countable product of two
point space $\{0,1\}$ with product topology, with the two point
space $\{0,1\}$ being given the discrete topology and
$\mathcal{B}$ its Borel $\sigma$-algebra.

If we drop from above $X$ the countable set of those sequences of
zeros and ones which have only finitely many zeros or finitely
many ones, then the remaining set, say, $Y$, can be mapped one-one
into $[0,1)$ by the map $\xi(x_1,x_2,...)=\sum_{i=1}^{\infty}
{x_i}/{2^i}$.  The image of $Y$ under this map is $[0,1)-D$ where
$D$ is the set of rational numbers of the form $k/2^n$, $0\leq
k\leq 2^n$, $n\in\mathbb{N}$.

Now, if $x=(x_1, x_2, ...)\in Y$ and if $k$ is the first integer
s.t.\ $x_k=0$, then let us define the map, say, $V=\xi^{-\,1}T\xi$
as $Vx=(0,0,...,0,1,x_{k+1},x_{k+2},...)$. Then, $V$ replaces all
the ones up to the first zero by zeros and replaces the first zero
by one, leaving all other coordinates of $x$ unchanged.

We call the map $V$ on $Y$ the {\em Diadic Adding Machine\/} (DAM)
or the {\em Odometer}.

Now, a measure preserving automorphism $T$ on a probability space
$(X, \mathcal{B}, m)$ is ergodic iff $\forall\;A, B\in
\mathcal{B}$, $\frac{1}{n} \sum_{k=0}^{n-1}m\left( A\bigcap T^kB
\right) \to m\left( A\bigcap B\right)$ as $n\to\infty$. There are
two properties stronger than ergodicity which are also relevant to
us.

A measure preserving automorphism $T$ on a probability space $(X,
\mathcal{B}, m)$ is said to be {\em weakly mixing\/} iff
$\forall\;A, B\in \mathcal{B}$, $\frac{1}{n}
\sum_{k=0}^{n-1}|m\left( A\bigcap T^kB \right)-m\left( A\bigcap B
\right)| \to 0$ as $n\to\infty$. A measuring preserving
automorphism $T$ on $(X,\mathcal{B},m)$ is said to be {\em
mixing\/} if $\forall\;A, B \in \mathcal{B}$, $m\left( A\bigcap
T^kB \right) \to m\left(A\bigcap B\right)$ as $n\to\infty$.

If a measure preserving Borel automorphism $T$ is mixing then it
is weakly mixing, and if $T$ is weakly mixing then it is ergodic.
However, a ergodic $T$ need not be weakly mixing and mixing. Also,
an ergodic and weakly mixing automorphism $T$ need not be mixing.

Let $T_1$ be a measure preserving Borel automorphisms on a
probability space $(X_1,\mathcal{B}_1, m_1)$ and $T_2$ be that on
$(X_2,\mathcal{B}_2,m_2)$. We say that $T_1$ and $T_2$ are {\em
metrically isomorphic\/} if $\exists$ $X'_1\subseteq  X_1$ with
$m_1\left(X_1-X_1'\right)=0$, $X_2'\subseteq X_2$ with $m_2\left(
X_2-X_2'\right)=0$ and an invertible, {\em ie}, a one-one, onto,
measurable map with measurable inverse, measure preserving map
$\phi:X_1'\to X_2'$ s.t.\ $\phi T_1\phi^{-\,1}=T_2$.

A measure preserving automorphism $T$ gives rise to a {\em Unitary
Operator}, $U_{_T}$, as: $U_{_T}f=f\circ T,\;f\in
L^2(X,\mathcal{B},m)$. The unitary operator is linear, invertible
with $U_{_T}^{-\,1}f=f\circ T^{-\,1}$ and $L^2$-norm preserving,
{\em ie}, $||U_{_T}f||_{_2} =||f||_{_2}$.

We say that $\lambda$ is an {\em eigenvalue\/} of $U_{_T}$ if
$\exists$ a non-zero $f\in L^2(X,\mathcal{B},m)$, s.t.\ $f\circ
T=\lambda f$. Then,$f$ an {\em eigenfunction\/} with eigenvalue
$\lambda$. An eigenvalue is {\em simple}, if up to a
multiplicative constant, it admits only one eigenfunction.

Let $L_o^2(X,\mathcal{B},m)\,=\,\{ f \in L^2(X, \mathcal{B},
m)\;|\; \int\,f dm$ $=0\}$, the subspace of functions orthogonal
to the constant functions. It is $U_{_T}$-invariant.

Now, $1$ is always an eigenvalue of $U_{_T}$ and that $1$ is a
simple eigenvalue of $U_{_T}$ iff $T$ is ergodic. Further, since
$U_{_T}$ is unitary, all eigenvalues of $U_{_T}$ are of absolute
value one.

Then, weakly mixing automorphisms $T$ are precisely those for
which $U_{_T}$ has no eigenvalue other than $1$. Also, $T$ is
ergodic iff $1$ is not an eigenvalue of $U_{_T}$ on $L_o^2(X,
\mathcal{B}, m)$.

If $U_{_T}$ and $U_{_{T'}}$ are unitarily equivalent, $T$ and $T'$
are {\em spectrally isomorphic}. If measure preserving $T$ and
$T'$ are metrically isomorphic, then $U_{_T}$ and $U_{_{T'}}$ are
unitarily equivalent.

A measure preserving automorphism $T$ on a SPS $(X,
\mathcal{B},m)$ has {\em discrete spectrum\/} if $U_{_T}$ admits a
complete set of eigenfunctions. Then, if $T_1$ and $T_2$ are
spectrally isomorphic and $T_1$ has a discrete spectrum, then
$T_2$ also has a discrete spectrum and the corresponding unitary
operators have the same set of eigenvalues.

But, spectrally isomorphic measure preserving automorphisms are
not necessarily metrically isomorphic, in general. However, if the
measure preserving automorphisms defined on a SPS are ergodic with
discrete spectrum and are admitting the same set of eigenvalues,
then such spectrally isomorphic measure preserving automorphisms
are metrically isomorphic.

Note that in the case of a SPS, $U_{_T}$ can have at most a
countable number of eigenvalues, all of absolute value one.
Furthermore, in the same case, the eigenvalues of $U_{_T}$ form a
subgroup of the circle group $S^1$. Also, for each eigenvalue
$\lambda$ we can choose an eigenfunction $f_{\lambda}$ of absolute
value one so as to have $f_{\lambda}.f_{\nu}=f_{\lambda\nu}$ a.e.

Any two B-shifts are spectrally isomorphic but, in general, any
two B-shifts are not metrically isomorphic. Any two K-shifts are
spectrally isomorphic but, in general, any two K-shifts are not
metrically isomorphic.

For a finite partition $\mathcal{P}=\{P_1, P_2, ..., P_k\}$ of $X$
by members of $\mathcal{B}$, we define the {\em entropy\/} of
$\mathcal{P}$ to be $\sum\,-\;m(P_i)\,\log_e{m(P_i)}$ and denote
it by $H(\mathcal{P})$. Then, we have the {\em entropy of
$\mathcal{P}$ relative to an automorphism $T$\/} defined as:
$h(\mathcal{P}, T) = \mathrm{lim\;sup}
\frac{1}{n}\,H\left(\bigvee_{k=0}^{n-1}T^{-\,1}
\mathcal{P}\right)$, where $\bigvee_{k=0}^{n-1}T^{-\,1}
\mathcal{P}$ is used to denote the partition generated by
$T^{-\,1} \mathcal{P}$, $k=0$, ..., $n-1$. Note that the
$\mathrm{lim\;sup}$ is indeed an increasing limit.

Then, we have the {\em entropy of the automorphism $T$}, denoted
as $h(T)$, defined as: $h(T)=\mathrm{sup}\;h(\mathcal{P}, T)$,
where the supremum is taken over all finite partitions
$\mathcal{P}$ of $X$. Note that $h(T)$ is an invariant of the
metric isomorphism.

Then, if $T$ is a B-shift with independent generating partition
$\mathcal{P}=\{P_1, P_2, ...\}$ then its entropy is
$h(T)=\sum\;-\;m(P_i)\log_e{m(P_i)}$. Now, any two B-shifts with
the same entropy can be shown to be metrically isomorphic.

For any set $A\in \mathcal{B}$, the set
$\bigcup_{k=-\infty}^{\infty} T^kA$ is called as the {\em
saturation of A wrt $T$\/} or simply the {\em $T$-saturation of
$A$}. We denote it by {\it s}$_{_T}(A)$. A point $x\in A$ is said
to be a {\em recurrent point in $A$\/} if $T^nx$ returns to $A$
fimp $n$ and fimn $n$.

By Poincar\'{e} Recurrence Lemma, we can write $A$ as a disjoint
union of two measurable sets $B$ and $M$ s.t.\ every point of $B$
is recurrent in $B$ (hence also in $A$) and no point of $M$ is
recurrent so that $M\in\mathcal{W}_{_T}$. Clearly, it follows that
$\bigcup_{n=0}^{\infty} T^nB=\bigcup_{n=-\infty}^{\infty}T^nB=$
{\it s}$_{_T}(B)$, since every point of $B$ is recurrent in $B$.

At this point, we note that the P-sets partition the Einstein
space $\mathbb{B}$, the partition being countably infinite.
Further, each P-set is a measurable subset of $\mathbb{B}$ in
$\mathcal{B}_{\mathbb{B}}$. Then, we will be dealing with measure
preserving Borel automorphisms on the Einstein SPS $(\mathbb{B},
\mathcal{B}_{\mathbb{B}}, \mu)$ that are B-shifts.

Further, we also note that a P-set can be decomposed as a disjoint
union of two measurable sets of $\mathcal{B}$ as per the above
method based on the Poincar\'{e} Recurrence Lemma.

Now, given $x\in B$, let $n_{_B}(x)$ denote the smallest positive
integer s.t.\ $T^nx\in B$. Then, we can decompose $B$ into
pairwise disjoint sets $B_k,\;k\in \mathbb{N}$, where $B_k=\{x\in
B\; |\; n_{_B}(x)=k\}$ or, equivalently, $B_k=\{x\in B\;|\;
Tx\notin B, ..., T^{k-1}x \notin B, T^kx\in B\}$. Further, we have
$T^kB_k\subseteq B$ and that $B_k$, $TB_k$, ..., $T^{k-1}B_k$ are
pairwise disjoint.

Further, let $F_{\ell}=T^{\ell}\left( \bigcup_{k\,>\,\ell}B_k
\right)$ and note also that $F_{\ell} = TF_{\ell-1}-B$, where
$F_o=B$. Now, we have $\bigcup_{k=0}^{\infty}T^kB =
\bigcup_{k=0}^{\infty}\bigcup_{i=0}^{k-1}T^kB_k =
\bigcup_{k=0}^{\infty}F_k = \bigcup_{k=-\infty}^{\infty}T^kB =$
{\it s}$_{_T}(B)$, with the middle two unions being pairwise
disjoint unions.

We call the set $B$ the {\em base\/} and the union
$\bigcup_{k=1}^{\infty}T^{k-1}B_k$ the {\em top of the
construction}. The above construction is called as the {\em
Kakutani tower over base $B$}.

Now, if $m$ is any $T$-invariant probability measure on
$\mathcal{B}$ and if {\em we write
$B_{\star}=\bigcup_{k=0}^{\infty} T^kB$}, then we have $m\left(
B_{\star}\right) = \sum_{k=1}^{\infty} \sum_{i=0}^{k-1}m\left(
T^iB_k\right) = \sum_{k=1}^{\infty} k\, m\left(B_k\right) =
\int_B\,n_{_B}(x)\,dm$.

Let $m(B)\ge 0$. Then, we call the quantity
$\frac{1}{m(B)}\int_B\,n_{_B} (x)\,dm=m(B_{\star})/m(B)$ as the
{\em mean recurrence time of $B$}. Recall $A=B\bigcup M$, $M\in
\mathcal{W}_{_T}$. Then, $m(M)=m(M_{\star})=0$. Hence, $m(A)=m(B)$
and $m(A_{\star})=m(B_{\star})$. Thus, the above is also the {\em
mean recurrence time of $A$}.

Clearly, if $T$ is ergodic and $m(B)>0$ then we have
$B_{\star}=X\; (\mathrm{mod}\;m)$ since it is $T$-invariant and of
positive measure.

Now, consider the transformation $\mathcal{S}$ defined over
$B_{\star}$ as: \begin{eqnarray} \mathcal{S}(x)= \left\{ \n
\begin{array}{cccc}
T(x) &{\rm if} \;x&\notin\bigcup_{k=1}^{\infty}T^{k-1}B_k = {\rm
Top}\\ \\ T^{-\,k+1}(x) &{\rm if} \;x&\in T^{k-1}B_k, \,k=1,2,...
\end{array}\right.
\end{eqnarray}

\noindent Then, $\mathcal{S}$ is periodic, the period being $k$
for points in $B_k$, and $\mathcal{S}$ agrees with $T$ everywhere
except at the top of the Kakutani tower. Further, if
$B_{\star}=X$, then $\mathcal{S}$ is defined on all of $X$.

Now, suppose $C_1\supseteq C_2\supseteq C_3\supseteq ...$ is a
sequence of sets in $\mathcal{B}$ decreasing to an empty set and
s.t.\ $\forall\;n$, we have \begin{description} \item{(i)} every
point of $C_n$ is recurrent, and that \item{(ii)}
$\bigcup_{k=0}^{\infty}T^kC_n=X$. \end{description} Let
$\mathcal{S}_n$ be the periodic automorphism as defined above with
$B=C_n$. Then, $\forall\;n$, $\mathcal{S}_n$ and
$\mathcal{S}_{n+1}$ agree except on the Top $T_{n+1}$ of the
Kakutani tower whose base is $C_{n+1}$. But $T_n\supseteq T_{n+1}$
and since $C_n$ decreases to $\emptyset$, $T_n$ also decreases to
$\emptyset$. Then, given any $x$, $\exists\;n(x)$ s.t.\ $\forall\;
k\geq n(x)$, $\mathcal{S}_k(x)$ are all the same and equal to
$T(x)$. Thus, $T$ is a limit in this sense of the sequence of
periodic automorphisms. Hence, we obtain the {\em periodic
approximation of automorphism $\;T$}.

A very useful result, Rokhlin's Lemma, states that: If $T$ is
ergodic wrt the $\sigma$-ideal of null sets of a finite measure
$m$, then given $\epsilon>0$ and $n\in\mathbb{N}$, $\exists$ a set
$C$ s.t.\ $C$, $TC$, ..., $T^{n-1}C$ are pairwise disjoint and
$m\left( X- \bigcup_{k=o}^{n-1}T^kC\right)\;<\; \epsilon$.

Now, let $B\in\mathcal{B}$ be s.t.\ every point of $B$ is
recurrent. Following Kakutani, the {\em induced automorphism on
$B$} (mod $\mathcal{W}_{_T}$), denoted as $T_{_B}$, is then
defined as: $T_{_B}(x)=T^n(x), \;x\in B$, where $n=n_{_B}(x)$ is
the smallest positive integer for which $T^n(x)\in B$. Note that
$T_{_B}(x)=T^k(x)$ if $x\in B_k$, $k=1,2,3....$. Then, $T{_B}$ is
one-one, measurable and invertible with $T^{-\,1}(x)=T^n(x)$ where
$n$ is the largest negative integer s.t.\ $T^n(x)\in B$. Thus,
$T_{_B}$ is a Borel automorphism on $B$.

The induced Borel automorphism, $T_{_B}$, on $B$ has following
properties:
\begin{itemize} %(i)
\item $\mathrm{orb}(x,T_{_B})=B\bigcap\mathrm{orb}(x,T)$, $x\in B$
\item $T_{_B}$ is elementary iff $T$ restricted to {\it s}$_{_T}B$
is elementary
\item $W\subseteq B$ is $T_{_B}$-wandering iff $W$ is $T$-wandering
\item $\mathcal{W}_{_{T_{_B}}} =\mathcal{W}_{_T} \bigcap B$
\item if $T$ is ergodic and preserving a finite measure $m$ then
$T_{_B}$ is ergodic and preserves $m$ restricted to $B$,
\item If $\mathcal{N}$ is a $\sigma$-ideal in $\mathcal{B}$,
$\mathcal{W}_{_T}\subseteq \mathcal{B}$, and if $T$ is ergodic wrt
$\mathcal{N}$, then $T_{_B}$ is ergodic wrt the restriction of
$\mathcal{N}$ to $B$. In particular, if $T$ is ergodic wrt a
finite continuous measure $m$ then $T_{_B}$ is ergodic wrt the
restriction of $m$ to $B$
\item if $C\subseteq B$, then a point of $C$ is recurrent wrt $T$
iff it is recurrent wrt $T_{_B}$. If every point of $C$ is
recurrent then we have $T_{_C}=(T_{_B})_{_C}$.\end{itemize}

A broadened view of the induced automorphism defines it on a set
$A\in \mathcal{B}$ even if not every point of $A$ is recurrent.
For this, let us consider a set $B=\{x\in A\;|\; x
\;\mathrm{is\;recurrent \;in}\;A\}$. By Poincar\'{e} Recurrence
Lemma, $A-B\in\mathcal{W}_{_T}$ and every point of $B$ is
recurrent in $B$. Then, the broadened induced automorphism
$T_{_A}$ is defined on all of $A$ iff every point of $A$ is
recurrent; otherwise $T_{_A}$ is defined on $A\;(\mathrm{mod}\;
\mathcal{W}_{_T})$. All the earlier properties of the induced
automorphism remain valid $(\mathrm{mod}\;\mathrm{W}_{_T})$ under
this broadened definition of $T_{_A}$. Note however that the
stricter point of view is necessary for the descriptive aspects.

Now, consider a Borel automorphism $T$ on $(X, \mathcal{B})$ and
let $f$ be a non-negative integer-valued measurable function on
$X$.

Let $B_{k+1}=\{x \;|\;f(x)=k\}$, $k=0,1,2,...$, $C_k=
\bigcup_{\ell\,>\,k}B_{\ell}$, $F_k=C_k\times\{k\}$, $Y=
\bigcup_{k=0}^{\infty}F_k$. If $Z=X\times\{0,1,2,...\}$, then
$Y\subseteq Z$ is the set $Y=\{(x,n)\;\;0\geq n\geq f(x)\}=$
Points in $Z$ below and including the graph of $f$.

Define $\Lambda$ on $Y$ as:
\begin{eqnarray} \Lambda(k,j)= \left\{ \n
\begin{array}{cccc}
(b,j+1)&{\rm if}\;b\in B_k\; {\rm and}\; 0\leq j\leq k-1\\
\\ (\Lambda(b), 0) &{\rm if}\;b\in B_k\; {\rm and}\; j=k-1
\end{array}\right.
\end{eqnarray}
This $\Lambda$ is a Borel automorphism on the space $Y$. We call
it the {\em automorphism built under the function $\;f$\/} on the
space $X$. We call $X$ the {\em base space of $\;\Lambda$\/} and
$f$ the {\em ceiling function of $\Lambda$}. Note that if we
identify $X$ with $X\times \{0\}$, then $\Lambda_{_X}=T$ and we
write $\Lambda=T^f$.

The automorphism built under a function has the following
properties: \begin{itemize}\item If $B\in\mathcal{B}$ with every
point of $B$ being recurrent and $B_{\star}=X$, then $T$ is
isomorphic to $(T_{_B})^f$, where $f(x)=n_{_B}(x)$, \item If
$A\subseteq Y$ is the graph of a measurable function $\xi$ on $X$,
then $(T^f)_{_A}$ and $T$ are isomorphic by $x\mapsto(x,\xi(x))$.
In particular, $(T^f)_{_A}$ and $T$ are isomorphic when $A=$ graph
of $f$, \item If $A\subseteq Y$ is measurable then we can find a
measurable $B$ with the same saturation as $A$ under $T^f$ and
s.t.\ $\forall\;x\in X$, $B\bigcap \{(x,i)\;|\;0\leq i \leq
f(x)\}$ is at most a singleton. Indeed, $B=\{(x,i)\in
A\,|\,(x,j)\notin A,\, 0\leq j<i\}$ can be chosen, \item Given
$T^f$ and $T^g$, they are isomorphic to automorphisms induced by
$T^{f+g}$ on suitable subsets. If $Y_1=\{(x,i)\,|\,0\leq i\leq
f(x)+g(x)\}$ on which $T^{f+g}$ is defined, then the sets
$\{(x,i)\,|\,0\leq i\leq f(x)\}$ and $\{(x,i)\,|\,0\leq i\leq
g(x)\}$ are subsets of $Y_1$ on which $T^{f+g}$ induces
automorphisms which are isomorphic to $T^f$ and $T^g$
respectively, \item If $m$ is a $\sigma$-finite $T$-invariant
measure on $X$, then $\exists$ a unique $\sigma$-finite
$T^f$-invariant measure $m_{_Y}$ on $Y$ s.t.\ $m_{_Y}$ restricted
to $X\times\{0\}$ is $m$. The measure $m_{_Y}$ is finite iff
$m(X)$ is finite and $\int\,f\,dm$ is finite. Then, we have
$m_{_Y}(Y)=\sum_{k=1}^{\infty} k\,m(B_{k+1})=\int\, f\,dm <
\infty$. \item $T^f$ is elementary iff $T$ is elementary.
\end{itemize}

Now, given two Borel automorphisms $T_1$ and $T_2$, we say that
$T_1$ is a {\em derivative\/} of $T_2$, and write $T_1 \prec T_2$,
if $T_1$ is isomorphic to $(T_1)_{_A}$ for some $A\in\mathcal{B}$
with $\bigcup_{k=0}^{\infty}T^k_1A=X$. If $T_1$ is a derivative of
$T_2$, we call $T_2$ the {\em primitive of $T_1$}. Two Borel
automorphisms are said have a {\em common derivative\/} if they
admit derivatives which are isomorphic. Similarly, two
automorphisms are said to have a {\em common primitive\/} if they
admit primitives which are isomorphic. If $T_1 \prec T_2$, then
clearly $T_2=T_1^f$ for some $f$.

Then, a lemma due to von Neumann states that: Two Borel
automorphisms have a common derivative iff they have a common
primitive.

Now, we say that two Borel automorphisms $T_1$ and $T_2$ are {\em
Kakutani equivalent}, and we write $T_1 \kakueq T_2$, if $T_1$ and
$T_2$ have a common primitive, or, equivalently the automorphisms
T$_1$ and $T_2$ have a common primitive. The Kakutani equivalence
is reflexive, symmetric and transitive. Therefore, the Kakutani
equivalence is an equivalence relation for Borel automorphisms.

Suppose $\mathcal{N}$ is a $\sigma$-ideal in $\mathcal{B}$. Then,
we say that $T_1$ and $T_2$ are {\em Kakutani equivalent
$(\mathrm{mod}\; \mathcal{N})$\/} if we can find two sets $M, N\in
\mathcal{N}$, $M$ being $T_1$-invariant and $N$ being
$T_2$-invariant, s.t.\ $T_1|_{_{X-M}} \kakueq T_2|_{_{X-N}}$. When
$\mathcal{N}$ is the $\sigma$-ideal of $m$-null sets of a
probability measure $m$ invariant under $T_1$ and $T_2$ both, we
get the measure theoretic Kakutani equivalence of Borel
automorphisms \cite{orw}.

Given a Borel automorphism $T$, a system of pairwise disjoint sets
$(C_o, C_1, ..., C_n) \in \mathcal{B}$ is called a {\em column\/}
if $C_i=T^iC_o,\;0\leq i \leq n$. $C_o$ is called the {\em base of
the column\/} and $C_n$ is called the {\em top of the column}. If
$D_o\subseteq C_o$, then $(D_o, TD_o, ..., T^nD_o)$ is called a
{\em sub-column of $(C_o, ..., C_n)$}.

Two columns $(C_o, ..., C_n)$ and $(B_o, ..., B_m)$ are said to be
{\em disjoint\/} if $C_i\bigcap B_j=\emptyset\;\forall \;i\neq j$.
A finite or a countable system of pairwise disjoint columns is
called a {\em $T$-tower}.

A $T$-tower with $r$ pairwise distinct columns may be written as
$\{C_{ij}\;|\; 0\leq i\leq n(j),\; 1\leq j\leq r\}$ where
$\{C_{0j}, ..., C_{n(j)j}\}$ is its j-th column.

Sets $C_{ij}$ are {\em constituents of the $T$-tower},
$\bigcup_kC_{0k}$ is a {\em base of the $T$-tower\/} and
$\bigcup_kC_{n(k)k}$ is a {\em top of the $T$-tower}. The number
of distinct columns in a $T$-tower is a {\em rank of the
$T$-tower}.

A $T$-tower is said to {\em refine a $S$-tower\/} if every
constituent of $T$-tower is a subset of a constituent of the
$S$-tower.

$T$ has {\em rank at most $r$\/} if there is a sequence
$T_n,\;n\in \mathbb{N}$, of $T_n$-towers of rank $r$ or less s.t.\
$T_{n+1}$ refines $T_n$ and the collection of sets in $T_n$, taken
over all $n$, generates $\mathcal{B}$. Then, $T$ has rank $r$ if
$T$ has rank at most $r$ but does not have rank at most $r-1$. If
$T$ does not have rank $r$ for any finite $r$, then $T$ has
infinite rank.

Given a Borel automorphism $T$ on $(X,\mathcal{B})$, a partition
$\mathcal{P}$ of $X$, $\mathcal{P} \subseteq \mathcal{B}$, is a
{\em generator of $T$\/} if $\bigcup_{k=1}^{\infty}
T^k\mathcal{P}$ generates $\mathcal{B}$. A set $A\in\mathcal{B}$
is {\em decomposable $(\mathrm{mod}\;\mathcal{W}_{_T})$} if we can
write $A$ as a disjoint union of two Borel sets $C$ and $D$ s.t.\
{\it s}$_{_T}(C)=${\it s}$_{_T}(D)=${\it
s}$_{_T}(A\;(\mathrm{mod}\;\mathcal{W}_{_T}))$.

Let $\mathcal{P}=\{P_1, P_2,..., P_n\}\subseteq\mathcal{B}$ be a
partition of $X$ and let a measurable $C$ be s.t.\
$\bigcup_{k=0}^{\infty}T^kC=X$. Then, on the basis of the first
return time $n(x)$ of each $x\in C$ and pairwise disjoint sets
$E_i=\{ x\in C\;|\; n(x)=i\}$ with union $\bigcup_iE_i=C$, there
exists a countable partition of $\{D_1,D_2,...\}$ of $C$ s.t.\
each $P_i$ is a disjoint union of sets of the form $T^kD_i$,
$k=1,2,...$, $i=1,2,...$.

Now, a one-one and onto map $T:X\to X$ s.t.\ $T^kx\neq x$ for all
$k\neq 0$, and for all $x\in X$ is called a {\em free map}.

Every free Borel automorphism $T$ on a SBS $(X,\mathcal{B})$ is
\cite{djk} orbit equivalent to an induced automorphism by the DAM.

Further, every Borel set $A\in\mathcal{B}$ is clearly decomposable
$(\mathrm{mod}\;\mathcal{W}_{_T})$ for $T$ being a free Borel
automorphism on a countably generated and countably separated SBS.

Furthermore, given a free Borel automorphism $T$ on a countably
generated and countably separated SBS $(X,\mathcal{B})$, there
exists a sequence $C_n,\;n\in\mathbb{N}$, of Borel sets decreasing
to an empty set with {\it s}$_{_T}(C_n)=${\it s}$_{_T}(X-C_n)=X\;
\forall\;n$, s.t.\ $\forall\;n$ the sets $C_n$, $TC_n$, ...,
$T^{n-1}C_n$ are pairwise disjoint, and s.t.\
$\bigcap_{n=1}^{\infty}C_n=C_{\infty}$, say, is $T$-wandering.

Also, given a Borel automorphism $T$ on a countably generated and
countably separated SBS $(X,\mathcal{B})$, there exists a sequence
$T_n,\;n=1,2,...$ of periodic Borel automorphisms on $X$ s.t.\
$\forall\;x$, $Tx=T_n x$ for all sufficiently large $n$.

Hence, the descriptive version of Rokhlin's theorem \cite{rokhlin}
on generators is obtained \cite{weiss1} as: every free Borel
automorphism on a countably generated and countably separated SBS
admits a countable generator in a strict sense.

Note also that $T$ admits a countable generator iff $T$ admits at
most a countable number of periodic points \cite{kechris}.

Now, two subsets of $X$, $A, B \in {\cal B}$, are said to be {\em
equivalent by countable decomposition}, and we write $A \sim B$,
if
\begin{description}\item{(a)} $A=\bigcup_{i=1}^{\infty}A_i$, $A_i\bigcap
A_j=\emptyset$ for $i\neq j$, and $A_i \in {\cal B},\, i=1, 2,
...$ \item{(b)} $B=\bigcup_{i=1}^{\infty}B_i$, $B_i\bigcap
B_j=\emptyset$ for $i\neq j$, and $B_i \in {\cal B},\, i=1, 2,
...$ \item{(c)} there exist $n_1$, $n_2$, $... \in \mathbb{N}$
s.t.\ $\forall\;i\in\mathbb{N}, \;\;T^{n_i} A_i=B_i\;({\rm
mod}\,\mathcal{N})$.\end{description} The equivalence by countable
decomposition is an equivalence relation on ${\cal B}$.

Note that if $A_i\in {\cal B},\;i\in\mathbb{N}$ are pairwise
disjoint and $B_i\in {\cal B},\;i\in\mathbb{N}$ are pairwise
disjoint and if $\forall\; i\in\mathbb{N}, \; A_i\sim B_i$ then
$\bigcup_{i=1}^{\infty} A_i \;\sim\;\bigcup_{i=1}^{\infty}B_i$.

If $A\sim B$, then we say that $B$ is a {\em copy of $A$\/} and
then, $A$ and $B$ have the same measure wrt a $T$-invariant
$\sigma$-finite measure.

Further, we say that $A$ and $B$ are {\em equivalent by countable
decomposition $(\mathrm{mod}\;m)$}, and we write $A\sim B\;
(\mathrm{mod}\;m)$, if there exist sets $M$ and $N$ in
$\mathcal{B}$, of $m$-measure zero, s.t.\ $A\triangle M \sim
B\triangle N$.

A set $A \in {\cal B}$ is said to be {\em $T$-compressible in the
sense of Hopf\/} if there exists $B \subseteq A$ s.t\ $A\sim B$
and $m(A-B)>0$. Clearly, if the set $X$ is Hopf $T$-compressible
then every of its subsets $B\in{\cal B}$ is Hopf $T$-compressible.

If $\mu$ is a $T$-invariant finite measure on $\mathcal{B}$ and
having the same null sets as $m$, then $A\sim B\; (\mathrm{mod}
\;m)$ implies that $\mu(A)=\mu(B)$. Whenever such a $\mu$ exists,
no measurable sets of positive measure can be compressible in the
sense of Hopf and, in particular, $X$ is not Hopf
$T$-compressible.

In a descriptive setting, one can dispense with the measure and
consider only a SBS $(X,\mathcal{B})$ and a free Borel
automorphism $T$ on it.

Then, given $A, B\in \mathcal{B}$, we write $A\prec\prec B$ if
there exists a measurable subset $C\subseteq B$ s.t.\ $A \sim C$
and {\it s}$_{_T}(B-C)\;=\;${\it s}$_{_T}B$, which is the smallest
$T$-invariant set containing $B$.

Now, we say that $A$ is {\em $T$-compressible\/} if $A\prec\prec
A$ or, equivalently, if we can write $A$ as a disjoint union of
two sets $C,\,D\in\mathcal{B}$ s.t.\ $A\sim C$, and {\it
s}$_{_T}(A)=$ {\it s}$_{_T}(C)=$ {\it s}$_{_T}(D)$. The sets $C$
and $D$ together with the automorphism $T$ which accomplishes
$A\sim C$ is called a {\em compression of $A$}.

If $X$ is $T$-compressible, we say that {\em $T$ is
compressible\/} or that {\em $T$ compresses $X$}.

The above notion of compressibility has the following properties:
\begin{itemize} \item If $A\in\mathcal{B}$ is $T$-compressible then any
superset of $A$ in $\mathcal{B}$ having the same saturation as $A$
is compressible. In particular, {\it s}$_{_T}(A)$ is
$T$-compressible whenever $A$ is $T$-compressible, \item Since $T$
is a free automorphism, each orbit is infinite and
$T$-compressible as also the saturation of any $T$-wandering set.
However, every $T$-compressible $T$-invariant set in $\mathcal{B}$
is {\em not\/} the saturation of a $T$-wandering set in
$\mathcal{B}$ except in special cases, \item A finite non-empty
set is not $T$-compressible nor is a set $A$ $T$-compressible if
the orbit of some point intersects $A$ in a finite non-empty set.
Further, if there exists a $T$-invariant probability measure on
$\mathcal{B}$, then no set of positive measure is
$T$-compressible. In particular, $X$ is not $T$-compressible in
this case, \item Clearly, a subset of a $T$-compressible set need
not be $T$-compressible, \item If $E\in\mathcal{B}$ is
$T$-invariant, $T$-compressible, and if $F\in\mathcal{B}$ is a
$T$-invariant subset of $E$, then $F$ is $T$-compressible. The
countable pairwise disjoint union of $T$-invariant,
$T$-compressible sets in $\mathcal{B}$ is $T$-compressible.
Clearly, any countable union of $T$-invariant, $T$-compressible
sets in $\mathcal{B}$ is $T$-compressible, \item $T$-compressible
sets in $\mathcal{B}$ do not form a $\sigma$-ideal in
$\mathcal{B}$. However, $T$-invariant, $T$-compressible sets in
$\mathcal{B}$ are closed under countable union and taking of
$T$-invariant subsets in $\mathcal{B}$. Hence, the collection
$\mathcal{H}$ of subsets in $\mathcal{B}$ whose saturations are
$T$-compressible forms a $\sigma$-ideal in $\mathcal{B}$ and we
call $\mathcal{H}$ the {\em Hopf ideal}. \item
$\mathcal{W}_{_T}=\mathcal{H}$ iff $X\in \mathcal{W}_{_T}$.
\end{itemize}

Note that the Hopf ideal is also equal to the $\sigma$-ideal
generated by $T$-compressible sets in $\mathcal{B}$. Note that
$\mathcal{W}_{_T}\subseteq \mathcal{H}$ since the saturation of
every $T$-wandering set in $\mathcal{W}_{_T}$ is $T$-compressible,

Let $\mathcal{N}\subseteq \mathcal{B}$ be a $\sigma$-ideal s.t.\
\begin{description} \item{(1)} $T\mathcal{N}=T^{-\,1}\mathcal{N}=\mathcal{N}$ and
\item{(2)} $\mathcal{W}_{_T}\subseteq\mathcal{N}$.\end{description} The Hopf ideal
$\mathcal{H}$; the $\sigma$-ideal of $m$-null sets in
$\mathcal{B}$ for any $T$-invariant $\sigma$-finite measure on
$\mathcal{B}$; and the $\sigma$-ideal of $m$-mull sets when $T$ is
$m$-conservative are few such ideals.

Then, two sets $A,\,B\in\mathcal{B}$ are said to be {\em
equivalent by countable decomposition $(\mathrm{mod}\;
\mathcal{N})$\/} if we can find sets $M,\,N\in\mathcal{N}$ s.t.\
$A\triangle M\;\sim\;B\triangle N$. We then write $A\sim B\;
(\mathrm{mod}\; \mathcal{N})$. Note that if $A\sim B\;
(\mathrm{mod}\; \mathcal{N})$ then {\it s}$_{_T}(A)=$ {\it
s}$_{_T}(B)\;(\mathrm{mod}\; \mathcal{N})$. We write $A\prec\prec
B\;(\mathrm{mod}\; \mathcal{N})$ if there exists a set
$N\in\mathcal{N}$ s.t.\ $A\triangle N \prec\prec B\triangle N$.

A set $A$ is {\em compressible $(\mathrm{mod}\; \mathcal{N})$\/}
if $\exists\;N\in \mathcal{N}$ s.t.\ $A\triangle N$ is
$T$-compressible. For a $T$-invariant set in $\mathcal{B}$ all the
three notions of compressibility, namely, $T$-compressibility,
compressibility $(\mathrm{mod}\;\mathcal{W}_{_T})$ and
compressibility $(\mathrm{mod}\;\mathcal{H})$, are equivalent.

Now, suppose that $A,B\in\mathcal{B}$ are equivalent by countable
decomposition. Let $A=\bigcup_{i=1}^{\infty}A_i$,
$B=\bigcup_{i=1}^{\infty}B_i$ be pairwise disjoint partitions of
$A$ and $B$ respectively, s.t.\ for suitable integers $n_i,\;i\in
\mathbb{N}$, $T^{n_i}A_i=B_i$.

The map $S:A\to B$ defined by $S(x)=T^{n_i}x$ if $x\in A_i$ is an
{\em orbit preserving isomorphism\/} between $A$ and $B$. In case
$A$ and $B$ are equivalent by countable decomposition
$(\mathrm{mod}\;\mathcal{N})$ then $S$ will be defined between $A
\triangle N$ and $B\triangle M$ for suitable sets $M,n\in
\mathcal{N}$. Such a $S$ is an {\em orbit preserving
isomorphism\/} between $A$ and $B$ $(\mathrm{mod}\;\mathcal{N})$.

The following results are then easily obtainable for $A,B,C, D \in
\mathcal{B}$: \begin{description}\item{(a)} If $A\supseteq
B\supseteq C$ and $A\sim C$ then $A\sim B$ \item{(b)} If $A\sim C
\subseteq B$ and $B\sim D\subseteq A$ then $A\sim B$, \item{(c)}
If $A\supseteq B\supseteq C \; (\mathrm{mod}\;\mathcal{N})$ and
$A\sim C\;(\mathrm{mod}\; \mathcal{N})$, then $A\sim
B\;(\mathrm{mod}\;\mathcal{N})$, \item{(d)} If $A\sim
C\;(\mathrm{mod}\;\mathcal{N})$, $C\subseteq B \;
(\mathrm{mod}\;\mathcal{N})$, and $B\sim D\; (\mathrm{mod}\;
\mathcal{N})$, $D\subseteq A\;(\mathrm{mod}\;\mathcal{N})$, then
$A\sim B\;(\mathrm{mod}\;\mathcal{N})$.\end{description} Note that
for (c) and (d) we remove suitable sets in $\mathcal{N}$ from
$A,B,C,D$.

Now, a set $A\in\mathcal{B}$ is {\em incompressible\/} if it is
not compressible and it is {\em incompressible
$(\mathrm{mod}\;\mathcal{N})$\/} if it is not compressible
$(\mathrm{mod}\;\mathcal{N})$. Note however that $A\in\mathcal{B}$
is incompressible $(\mathrm{mod}\;\mathcal{N})$ does not mean that
$A\triangle N$ is incompressible for a suitable set $N\in
\mathcal{N}$. Note also that for a set in $\mathcal{B}$ to be
incompressible $(\mathrm{mod}\;\mathcal{N})$ it is sufficient that
its saturation is incompressible $(\mathrm{mod}\;\mathcal{N})$.

Let $N$ be a positive integer. Then, it is easy to see that there
exists $B\in \mathcal{B}$ s.t.\ {\it s}$_{_T}(B)=X$ and
$\forall\;x\in B$, its first return time, $n_{_B}(x)$, is s.t.\
$N\leq n_{_B}(x)\leq 2N$.

For any $F\in\mathcal{B}$ and $x\in X$, let us now define
$r_{\star}\left( x,F \right)=\liminf_{n\to\infty}\frac{1}{n}
\sum_{k=1}^n\mathbf{1}_F (T^kx)$ and $r^{\star}\left( x,F
\right)=\limsup_{n\to\infty} \frac{1}{n} \sum_{k=1}^n\mathbf{1}_F
(T^kx)$ where $\mathbf{1}_F$ is the identity function on set $F$.

Then, we note that given $0\leq b \leq 1$ and $\epsilon >0$, there
exists $F\in\mathcal{B}$ s.t.\ {\it s}$_{_T}(F)=X$ and
$b-\epsilon<\;r_{\star}(x,F)$, $r^{\star}(x,F)<b+\epsilon$.

We also note that, if $0<b<1$, then there exists $F\in\mathcal{B}$
s.t.\ {\it s}$_{_T}(F)=$ {\it s}$_{_T}(X-F)$ and for all $x\in X$,
$0< r^{\star}(x,F)<b$.

Further, for any $F\in\mathcal{B}$ and $\epsilon>0$, there exists
a measurable $G\subseteq F$ s.t.\ {\it s}$_{_T}(G)=$ {\it
s}$_{_T}(F-G)=$ {\it s}$_{_T}(F)$ and $r^{\star}(x,G)<\epsilon\;
(\mathrm{mod}\;\mathcal{W}_{_T})$.

Now, we have a key dichotomy: Let $E,F\in\mathcal{B}$ and let $f=
\mathbf{1}_{_E}-\mathbf{1}_{_F}$. Then, there exists a
$T$-invariant set $N\in\mathcal{W}_{_T}$ s.t.\ if $x\in X-N$, then
either \begin{description}\item{(a)}
$\forall\;y\in\;\mathrm{orb}(x,T)$, there exists $n\geq 0$ s.t.\
$\sum_{k=0}^nf(T^ky)\geq 0$, \\\centerline{Or} \item{(b)} the set
of $y\in\; \mathrm{orb}(x,T)$ s.t.\ $\forall\;n\geq 0$,
$\sum_{k=0}^n f(T^ky) < 0$ is unbounded to the left and right.
\end{description} These are {\em mutually exclusive\/} conditions.

Furthermore,  consider any decomposition of $X$ into pairwise
disjoint $T$-invariant sets $X_o$, $X_1$ $X_2$, $N$ with
$N\in\mathcal{N}$ and $X_o$, $X_1$ $X_2$ satisfying the properties
\begin{description}\item{(c)} $E\bigcap X_1 \prec\prec F\bigcap X_1$,
\item{(d)} $E\bigcap X_o\;\sim\; F\bigcap X_o$, \item{(e)} $E\bigcap X_2
\prec\prec F\bigcap X_2$.\end{description}

Such a decomposition will have the properties that \begin{itemize}
\item for $x\in X_1\;(\mathrm{mod}\;\mathcal{H})$ the set, say,
$A(x)=\{y\in\;\mathrm{orb}(x,T)\;|\; \sum_{k=0}^nf(T^ny)\,>
\,0\;\forall \;n\geq 0\}$ is unbounded to left and right, \item
for any $x\in X_o\;(\mathrm{mod}\;\mathcal{H})$, for all
$y\in\;\mathrm{orb}(x,T)$ $\exists\;n\geq 0$ s.t.\
$\sum_{k=0}^nf(T^ky)=0$, \item for $x\in X_2(\mathrm{mod}\;
\mathcal{H})$, the set, say, $B(x)=\{y\in\;\mathrm{orb}(x,T)\;|\;
\sum_{k=0}^nf(T^ny)\,<\,0\;\forall \;n\geq 0\}$ is unbounded to
left and right. \end{itemize}

Moreover, $(\mathrm{mod}\;\mathcal{H})$, we have that \beq
\{x\;|\; r_{\star}(x,E)\,<\,r_{\star}(x,F)\} &\subseteq& X_2,\n \\
\{x\;|\; r^{\star}(x,E)\,<\,r^{\star}(x,F)\} &\subseteq& X_2,\n \\
\{x\;|\; r_{\star}(x,E)\,>\,r_{\star}(x,F)\} &\subseteq& X_1,\n \\
\{x\;|\; r^{\star}(x,E)\,>\,r^{\star}(x,F)\} &\subseteq& X_1\n
\eeq

Then, we have the following measure free version of the Birkhoff
point-wise Ergodic Theorem as: For any $E\in \mathcal{B}$, the set
of points $x$ for which limit $\lim_{n\to\infty}
\frac{1}{n}\sum_{k=0}^{n-1} \mathbf{1}_E(T^kx)$ does not exist
belongs to the Hopf ideal $\mathcal{H}$. That is to say, the set
$\{x\;|\;r_{\star}(x,E)\, <\,r^{\star}(x,E)\}$ is compressible.

For any $E\in \mathcal{B}$, let us now write $m(E,x)=\lim
\frac{1}{n} \sum_{k=0}^{n-1}\mathbf{1}_{_E}(T^kx)$. This $m$ is
countably additive $(\mathrm{mod}\;\mathcal{H})$ and
$T$-invariant. Moreover, we can show that
$m(E,x)=0\;(\mathrm{mod}\;\mathcal{H})$ iff $E\in \mathcal{H}$.

Now, let the Polish topology $\Gamma$ on $X$ possess a countable
clopen base $\mathcal{U}$ that is closed under complements, finite
unions and applications of $T$. There then exists a $T$-invariant
set $N\in\mathcal{H}$ s.t.\ $\forall\;x\in X-N$, $m(A\bigcup B, x)
= m(A,x)+m(B,x)$ whenever $A,B\in \mathcal{U}$ and $A\bigcap B =
\emptyset$.

Fix $x\in X-N$ and let us write $m(A,x)=m(A)$, $A\in\mathcal{U}$.
For any $B\subseteq X$, let us define $m^{\star}(B)=\inf\left\{
\sum_{i=1}^{\infty} m(U_i)\;|\;B\subseteq \bigcup_{i=1}^{\infty}
U_i\right.$, $\left.U_i\in\mathcal{U}\;\forall\;i\right\}$. This
$m^{\star}$, an outer measure on $\mathfrak{P}(X)$, is
$T$-invariant, bounded by one and $m^{\star}(X)=1$.

Recall \cite{halmos1} that an outer measure $\mu^{\star}$ on the
power set of a metric space $(X,d)$ is called a {\em metric outer
measure\/} if $\mu^{\star}(E\bigcup
F)=\mu^{\star}(E)+\mu^{\star}(F)$ whenever $d(E,F)>0$. If
$\mu^{\star}$ is a metric outer measure on $(X,d)$ then all open
sets, hence, all Borel sets, are $\mu^{\star}$-measurable. Then,
$m^{\star}$ defined above is a metric outer measure on $X$. The
restriction of $m^{\star}$ to $\mathcal{B}$ is a countably
additive $T$-invariant probability measure on $\mathcal{B}$.

Further, if $T$ is not free, then it has a periodic point on whose
orbit we can always put a $T$-invariant probability measure.

Therefore, if $T$ is a Borel automorphism (free or not) of a SBS
$(S,\mathcal{B})$ s.t.\ $X$ is $T$-incompressible, then there
exists a $T$-invariant probability measure on $\mathcal{B}$. This
is Hopf's theorem.

Now, a set $A\in\mathcal{B}$ is {\em weakly $T$-wandering\/} if
$T^nA$ are pairwise disjoint for $n$ in some infinite subset of
integers. Then, a non-singular automorphism $T$ on a probability
space $(X,\mathcal{B},m)$ admits \cite{hakaku} an equivalent
$T$-invariant probability measure iff there is no weakly
$T$-wandering set of positive measure.

But, $T$-compressibility of $X$ does not imply the existence of a
weakly $T$-wandering set $W\in \mathcal{B}$ s.t.\ {\it
s}$_{_T}(W)=X$ \cite{ehn}. If a measurable $A\in\mathcal{B}$ is
$T$-compressible then {\it s}$_{_T}(A)\prec\prec A$ and {\it
s}$_{_T}\sim A$.

Let $T_1$ and $T_2$ be Borel automorphisms on a SBS. Then, if
$T_1$ and $T_2$ are orbit equivalent and if $T_1$ has an orbit of
length $n$ then so has $T_2$ and vice versa. Moreover, the
cardinality of the set of orbits of length $n$ for $T_1$ and $T_2$
is the same. Further, if $\mathbf{c}_k(T_1)$ is the cardinality of
the set of orbits of length $k$, then for each $k\leq \aleph_o$,
$\mathbf{c}_k(T_1)=\mathbf{c}_k(T_2)$ whenever $T_1$ and $T_2$ are
orbit equivalent.

Dye's theorem \cite{dye} proves that: any two free ergodic measure
preserving Borel automorphisms on a SPS $(X,\mathcal{B},m)$ are
orbit-equivalent $(\mathrm{mod}\;m)$. Furthermore, we also note
that if $T_1$ and $T_2$ are Borel automorphisms both compressible
and not admitting Borel cross-sections, then $T_1$ and $T_2$ are
orbit-equivalent \cite{chaubemgn}.

Let $M(X)=M(X,\mathcal{B},m)$ be the group of all measure
preserving automorphisms on the space $(X,\mathcal{B},m)$. Two
automorphisms in $M$ are identified if they agree a.e.

For a $T\in M$, let us denote by $[T]$, called the {\em full group
of $T$}, the collection of all $\tau\in M$ s.t.\ f.a.e. $x\in X$,
$\tau(x)=T^n(x)$ for some integer $n=n(x)$. Note that $\tau\in
[T]$ iff $\mathrm{orb}(x,\tau)\subseteq \mathrm{orb}(x,T)$ f.a.e
$x\in X$, or equivalently, there exists a decomposition of $X=
\bigcup_{n\in\mathbb{Z}}A_n\;(\mathrm{mod}\;m)$ s.t.\ $X=
\bigcup_{n\in\mathbb{Z}}T^nA_n\;(\mathrm{mod}\;m)$, $T^nA_n$ being
pairwise disjoint, and $\tau(x)=T^n(x)$ for $x\in A_n,\;n\in
\mathbb{Z}$.

Let $A\in\mathcal{B}$ and $\tau\in[T]$. We shall write
$\tau\in[T]^+$ on $A$ in case $\tau(x)=T^n(x)$, where $n=n(x)>0$
a.e. on $A$.

An automorphism $T$ is called {\em set periodic with period $k$},
for some positive integer $k$, if there exists a partition
$\mathcal{P}=\{D_1, D_2, ..., D_k\}$ of $X$ associated with $T$
s.t.\ $D_i=T^{i-1}D_1$, for $1\leq i\leq k$ with each $D_i\in
\mathcal{B}$.

If every $x$ is $T$-periodic with period $k$, then it is clear
that $T$ is set periodic with period $k$. However, it should also
be noted that $T$ can be set periodic without having any periodic
points.

An automorphism $T\in M(X)$ is called a {\em weak von Neumann
automorphism\/} if \begin{description}\item{(1)} $T$ is set
periodic with period $2^n$ for every $n\in\mathbb{N}$, \item{(2)}
There exists a sequence $\{ \mathcal{D}_n(T)=\left( D_1^n, ...,
D_{2^n}^n\right)\}, \;n\in \mathbb{N}$, of partitions of $X$
associated with $T$ satisfying
\begin{description}\item{(a)} $D_i^n=D_i^{n+1}\bigcup D_{i+2^n}^{n+1}$,
for $i=1,2$, ..., $2^n$, $n\in\mathbb{N}$ \item{(b)}
$D_i^n=T^{i-1}D_i^n$, for $i=1,2$, ..., $2^n$,
$n\in\mathbb{N}$.\end{description} \end{description}

For $x\in D^n_1$, we call the finite sequence $(x, Tx, ...,
T^{2^n-1}x)$ a {\em fiber of length $2^n$}. Two points $u,v\in X$
are said to be {\em in the same fiber of length $2^n$\/} if for
some $x\in D_1^n$, $u=T^kx$, $v=T^{\ell}x$, where $0\leq k$, $\ell
< 2^n-1$.

If, in addition to the above (1) and (2), we have
\begin{description}\item{(3)} the $\sigma$-field generated by
$\bigcup_{n=1}^{\infty} \mathcal{D}_n (T)$ is equal to
$\mathcal{B}\;(\mathrm{mod}\;m)$,\end{description} $T$ is called
as a {\em von Neumann automorphism}.

This above condition (c) means that there exists a $T$-invariant
set $N\in\mathcal{B}$ which is $m$-null and s.t.\ the collection
$\{ D\bigcap (X-n)\;|\;D\in
\bigcup_{n=1}^{\infty}\mathcal{D}_n(T)\}$ generates the
$\sigma$-algebra $\mathcal{B}$ restricted to $X-N$, equivalently,
the sets $D_k^n$ taken over all $n$ and all $k$ separate the
points of $X-N$.

For a weak von Neumann automorphism $T$, let $\mathcal{P}_n(T)$
denote the algebra generated by $\mathcal{D}_n(T)$. Then,
$\mathcal{P}_n(T)\subseteq \mathcal{P}_{n+1}(T)$ and the union
$\mathcal{P}(T)=\bigcup_{n=1}^{\infty}\mathcal{P}_n(T)$ is again
an algebra. For $A\in\mathcal{B}$, write $d(A)=\inf \{ m(A
\triangle B)\;|\;B\in \mathcal{P}(T)\}$. If $d(A)=0$ for every $A$
in a countable collection which generates $\mathcal{B}$ then $T$
is a von Neumann automorphism.

A DAM or Odometer $V$ on $\{0,1\}^{\mathbb{N}}$ is a von Neumann
automorphism. Furthermore, any two von Neumann automorphisms are
isomorphic modulo $m$-null sets.

Now, for ergodic $T\in M(X)$ and $\forall\;\;A, B\in \mathcal{B}$
with $0< m(A)=m(B)$, there exists a $J\in [T]$ s.t.\ $JB=A$ and
$J\in[T]^+$ on $B$. Therefore, if $m(A)=m(B)$, then $T_{_A}$ and
$T_{_B}$ are orbit equivalent. Indeed, $J$ when viewed as an
isomorphism from $A$ to $B$ establishes orbit equivalence
$(\mathrm{mod}\;m)$ between $T_{_A}$ and $T_{_B}$.

Moreover, let $T\in M(X)$ be ergodic and let $\epsilon >0$ be
s.t.\ $\epsilon < m(X)$. Then, there exists $A\in \mathcal{B}$
s.t.\ $A\bigcap TA=\emptyset$ and $m(X-A\bigcup TA)=\epsilon$.
Also, there exists a weak von Neumann automorphism $\omega\in [T]$
s.t.\ $[\omega]=[T]$.

Further, if $\tau_1\in[T]$ is a set periodic automorphism with
period $2^{^K}$ s.t.\ $\mathcal{D}(\tau_1)=(D_1, ..., D_{2^K})$ is
a partition of $X$ associated with $\tau_1$, then, for any
$\epsilon>0$ and any set $A\in\mathcal{B}$, there exists a weak
von Neumann automorphism $\tau_1\in [T]$ and an integer $L>0$ that
satisfy \begin{description} \item{(a)} $[\tau_1]=[\tau_2]$
\item{(b)} $\mathcal{D}(\tau_1)\subseteq \mathcal{D}_n(\tau_2)$ for
all $n\geq L$, where $\{\mathcal{D}_n
(\tau_2)\;|\;n\in\mathbb{N}\}$ are the partitions of $X$
associated with $\tau_2$ \item{(c)} $\{x\;|\; \tau_2(x)\neq
\tau_1(x)\} \subseteq D_{2^K}\in \mathcal{D}(\tau_1)$ \item{(d)}
for $n\geq L$, we have $m(A-A'_n)< \epsilon$, $m(A''-A)<
\epsilon$, where $A'_n= \bigcup D$ where union is over
$\mathcal{D}'_n= \{D\in \mathcal{D}_n(\tau_2)\;|\; D\subseteq A\}$
and $A''_n= \bigcup D$ where the union is over
$\mathcal{D}''_n=\{D\in\mathcal{D}_n (\tau_2)\;|\; m(A\bigcap
D)>0\}$. \end{description}

Under the same hypotheses as above, if we have in addition that
$\tau_1\in[T]^+$ on $X-D_{2^K}$ for $D_{2^K}\in\mathcal{D}
(\tau_1)$, then the weak von Neumann automorphism $\tau_2\in[T]$
and the positive integer $L>0$ chosen above also satisfy
\begin{description}\item{(e)} $\tau_2\in[T]^+$ on $X-D_{2^L}^L$ for $D_{2^L}^L \in
\mathcal{D}_L (\tau_2)$.\end{description}

Furthermore, there exists an integer $P>L$, and $C\in \mathcal{B}$
with $m(C)<\epsilon$ s.t.\ the following holds:
\begin{description}\item{(f)} $C(x,Tx)$ does not intersect $D_{2^P}^P\in
\mathcal{D}_{_P}$ for all $x\in X-C$, where for $y\in
\mathrm{orb}(x,\tau_2)$ with $\tau_2^{n(x)}x=y$ and $C(x,y)=(x,
\tau_2x, ..., \tau_2^nx=y)$, if $n=n(x)\geq 0$ and $C(x,y)=(x,
\tau_2^{-\,1}x, ..., \tau_2^nx=y)$, if $n=n(x)<0$.
\end{description} In other words, $x$ and $Tx$ belong to the same
$\tau_2$-fiber of length $2^P$ for any $x\in X-C$.

Then, given a free ergodic measure preserving automorphism $T$ on
a SPS $(X,\mathcal{B}, m)$, there exist two von Neumann
automorphisms $\tau_1$ and $\tau_2$ in $[T]$ s.t.\ (i) $\tau_1\in
[T]^+$ on $X$ and (ii) $[\tau_1]=[\tau_2]$.

Note that when two Borel automorphisms on $(X,\mathcal{B})$ are
free and uniquely ergodic, then the orbit equivalence holds
without discarding any set of measure zero. Moreover, any two free
Borel automorphisms on $(X,\mathcal{B})$, each admitting $n$
invariant ergodic probability measures, are orbit equivalent
whether we have $n$ as finite or countable or uncountable
\cite{djk}.

Now, we note that Krieger \cite{krieger} introduces an invariant
called the {\em ratio set, $r(T)$, of automorphism $T$\/} as a
closed subset of $[0,\infty)$ and $r(T)\bigcap (0,\infty)$ is a
closed multiplicative subgroup of $(0,\infty)$. Then, if
$r(T)=r(\tau)=[0,\infty)$ or if $r(T)= r(\tau)=\{0\}\bigcup
\{\alpha^k\;|\;k\in\mathbb{Z}\}$ for some $\alpha,\;0<\alpha<1$,
then $T$ and $\tau$ are orbit equivalent $(\mathrm{mod}\;m)$.

Extending these concepts to more general group actions is
possible. Then, let $G$ be Polish group of Borel automorphisms
acting in a jointly measurable manner on a SBS $(X,\mathcal{B})$.
Then, if $X$ is incompressible wrt the $G$-action then there
exists a probability measure on $\mathcal{B}$ invariant under the
$G$-action \cite{trim6}.

However, note that further generalizations than above are limited
by counter examples.

For example, let $G$ now denote the group of {\em all\/} Borel
automorphisms of an uncountable Polish space with the property
that the set $\{ x\;|\; gx\neq x\}$ is of the first Baire
category. Then, $X$ is not compressible. (We note that the
Einstein space is of this type.)

A set of the first Baire category is also called as a {\em meagre
set}. Now, the $\sigma$-ideal $\mathcal{H}_{_G}$ generated by
$G$-compressible sets in $\mathcal{B}$ is the $\sigma$-ideal of
meagre Borel subsets of $X$. Hence, $X\notin \mathcal{H}_{_G}$.
However, every probability measure on $\mathcal{B}$ is supported
on a meagre set. Therefore, a $G$-invariant probability measure on
$\mathcal{B}$ does not exist.

A {\em flow\/} on a SBS $(X,\mathcal{B})$ is said to be {\em
non-singular\/} wrt a $\sigma$-finite measure $\mu$ on
$\mathcal{B}$ if $\mu(A)=0$ implies that $\mu(T_t(A))=0$ for all
$A\in\mathcal{B}$ and $t\in \mathbb{R}$. In case,
$\mu(T_t(A))=\mu(A)$ for all $t\in \mathbb{R}$ and
$A\in\mathcal{B}$, then we say that the {\em flow preserves
$\mu$}.

Consider topology $\Gamma$ on $X$ in place of the Borel structure
$\mathcal{B}$. If the map $(t,x)\mapsto T_tx$ is continuous on
$\mathbb{R}\times X$, then the flow is {\em jointly continuous\/}
or simply {\em continuous}. Here $\mathbb{R}$ is given the usual
topology and $\mathbb{R}\times X$ is given the product topology.
Note that each $T_t$ is a homeomorphism of $X$ onto itself. We
also call a jointly continuous flow a {\em flow of
homeomorphisms}. It is often called as a {\em one parameter group
of automorphisms}.

Let $\sigma$ be a Borel automorphism on a SBS $(Y,\mathcal{C})$
and let $f$ be a positive Borel function on $Y$ s.t.\ $\forall\;y
\in Y$, the sums $\sum_{k=0}^{\infty}f(\sigma^ky)$, $\sum_{k=0}
^{\infty} f(\sigma^{-\,k}y)$ are infinite. Let $X=\{(y,t)\;|\;
0\leq t < f(y)\}$. Then, $X$ is the subset of $Y\times\mathbb{R}$
strictly under the graph of $f$. Give $Y\times\mathbb{R}$ the
product Borel structure and restrict it to $X$. We then obtain a
new Borel space $(X,\mathcal{B})$.

A jointly measurable flow $T_t,\;t\in\mathbb{R}$, on $X$ can be
defined as follows: a point $(y,u)\in X$ moves vertically up with
``unit speed'' until it reaches the point $(y,f(y))$ when it goes
over to $(\sigma(y),0)$ and starts moving up again with unit
speed. The term ``unit speed'' means that the linear distance
travelled in time $t$ equals $t$. The point thus reached at time
$t>0$ is defined to be $T_t(y,u)$. For $t<0$, $T_(y,u)$ is defined
to be the point $(y',u')$ s.t.\ $T_{-\\,t}(y',u')-(y,u)$. The
point $(y,0)$ is called the {\em base point of $(y,u)$}.

Analytically, the above is expressible as follows: Let $x=(y,u)\in
X$, and let $t\geq 0$. Then, $T_t(x)= T_t(y,u)=
\left(\sigma^ny,t+u-\sum_{k=0}^{n-1}f(\sigma^ky)\right)$ where $n$
is the unique integer s.t.\ $\sum_{k=0}^{n-1}f(\sigma^ky)\leq t+u<
\sum_{k=0}^nf(\sigma^ky)$. If $t<0$, the expression is $T_t(x)=
\left(\sigma^{-\,n}y,t+u+\sum_{k=1}^{n}f(\sigma^{-\,k}y)\right)$
where $n$ is the unique integer s.t.\ $0\leq t+u+ \sum_{k=1}^n
f(\sigma^{-\,k}y)< f(\sigma^{-\,n}y)$. It is understood that
$\sum_{k=0}^{-\,1}$ and $\sum_{k=0}^0$ are equal to zero. It is
easy to verify that $T_t, \;t\in\mathbb{R}$ is indeed a flow on
$X$.

The flow $T_t, \;t\in\mathbb{R}$ as defined above is called the
{\em flow (or special flow) built under the function $f$ with base
automorphism $T$ and base space $(Y, \mathcal{C})$}.  Note that a
flow built under a function is a continuous version of
automorphism built under a positive integer-valued function. We
thus use the notation of $T^f$ for the continuous case also.

Let the base space $Y$ be Polish, the base automorphism $\sigma$ a
homeomorphism of $Y$ and $f$ continuous on $Y$. Let us give
$Y\times \mathbb{R}$ the product topology, where $\mathbb{R}$ has
the usual topology. Let $\bar{X}=\{(y,t)\;|\;0\leq t\leq f(y)\}$
be the closure of $X\subseteq Y\times\mathbb{R}$. Now, define $g:
\bar{X}\to X$ by $g(y,t)=(y,t)$ if $0\leq t\leq f(y)$ and $g(y,t)=
(\sigma y,o)$ if $t=f(t)$. The map $g$ identifies the point
$(f,f(y))$ with $(\sigma y, 0)$. Let $\Gamma$ be the largest
topology on $X$ that makes $g$ continuous. Under this topology,
the flow  $\sigma^f$ is a jointly continuous flow of
homeomorphisms on $X$. The topology $\Gamma$ can be shown to be a
Polish Topology.

Therefore, a jointly measurable flow is also jointly continuous
wrt a suitable complete separable metric topology, the Polish
topology, on $X$ which also generates the $\sigma$-algebra
$\mathcal{B}$.

Let $T_t,\;t\in\mathbb{R}$ be a jointly measurable flow (without
fixed points) on a SBS $(X,\mathcal{B})$. Suppose we are able to
choose on each orbit of $T_t$ a non-empty discrete set of points
s.t\ the totality of these points taken over all orbits is a Borel
set in $\mathcal{B}$. That is, we suppose that there exists a
Borel set $Y\subseteq X$ s.t.\ $\forall\;x\in X$, the set $\{
t\;:\; T_t(x)\in Y\}$ is a non-empty and discrete subset of
$\mathbb{R}$. Such a subset is called a {\em countable
cross-section of the flow}.

Given a countable cross-section $Y$, we can write $X$ as the union
of three Borel sets $I$, $J$, $K$ as: $I=\{ x\in X\;|\; \{t\;|\;
T_t(x) \in Y\} \;\mathrm{is\;bounded\;below}\}$, $J=\{ x\in X\;|\;
\{t\;|\; T_t(x) \in Y\} \;\mathrm{is\;bounded\;above}\}$, $K=X-I
\bigcup J$. Let $i(x)=\inf \{t\;|\;T_t(x)\in Y\}$ and $j(x)=\sup
\{t\;|\; T_t(x) \in Y\}$. Then $i$ and $j$ are measurable
functions, so that $I$ and $J$, hence, also $K$, are measurable
sets.

Let us write $S(x)=T_{i(x)}(x),\;x\in I$. Then, $S(T_t(x))=S(x)$
$\forall\;t\in\mathbb{R}$ since $i(T_t(x))=i(x)-t$. The function
$S:I\to I$ is again measurable, and constant on orbits. Thus, if
we restrict the flow to $I$ then the orbit space admits a Borel
cross-section, the image of $I$ under $S$ being the required Borel
cross-section. Similarly for $J$. Therefore, in the set $I\bigcup
J$, the flow is isomorphic to a flow built under a function.

Then, there exists a Borel set $Y\subseteq X$ s.t.\ $\forall
\;x\in X$ the set $\{t\;|\;T_tx\in Y\}$ is non-empty, countable,
and discrete in $\mathbb{R}$, the flow is isomorphic to a flow
built under a function.

Thus, we note that every jointly measurable flow (without fixed
points) on a SBS admits a countable cross-section.

Further, for a jointly measurable $T_t$, it can be shown
\cite{vmw} that there exists a set $B\in\mathcal{B}$ s.t.\
$\forall\;x\in X$ the sets $\{t\in\mathbb{R}\;|\; T_tx\in B\}$ and
$\{t\in\mathbb{R}\;|\;T_tx\notin B\}$ have positive Lebesgue
measure.

Then, it can further be shown \cite{vmw} that every jointly
measurable flow $T_t,\;t\in\mathbb{R}$ (without fixed points) on a
SBS $(X,\mathcal{B})$ admits a measurable subset $Y\subseteq X$
s.t.\ $\forall\;x\in X$ the set $\{ t\;|\;T_tx\in Y\}$ is
non-empty and discrete in $\mathbb{R}$. Therefore, we see that
every jointly measurable flow (without fixed points) on a SBS is
isomorphic to a flow built under a function.

For general finite measure preserving flows, this result was
proved in \cite{ambrose} while the refinement and adaptation of
that method to a descriptive setting can be found in \cite{vmw}.

As a corollary, every jointly measurable flow without fixed points
on a SBS $(X,\mathcal{B})$ is a flow of homeomorphisms under a
suitable Polish topology on $X$ which generates $\mathcal{B}$.

Furthermore, for a jointly measurable flow $T_t\;t\in\mathbb{R}$
(without fixed points) on a SBS $(X,\mathcal{B})$ and given $0\leq
\alpha\leq 1$, there exists $B\in\mathcal{B}$ s.t.\ $\forall\;x\in
X$ the orbit of $x$ spends the proportion $\alpha$ of time in $B$,
that is, $\forall\;x\in X$, $\frac{1}{N}\;\mathrm{Lebesgue\;
measure}\;\{ t\;|\; T_tx\in B,\,0\leq t < N\}\to \alpha$ as $N\to
\infty$.

Note also that, under suitable modifications of the definition of
flow built under a function, these results hold for jointly
measurable flows with fixed points as well. These will be
considered separately at a suitable stage, however, not in the
present preparatory paper.

Now, consider the notion of a flow built under a function in a
measure theoretic setting. Let $(Y,\mathcal{B}_{_Y})$ be a SBS
equipped with a Borel automorphism $\tau:Y\to Y$ and a
$\sigma$-finite measure $n$ quasi-invariant for $\tau$. [A measure
$n$ on $\mathcal{B}$ is called {\em quasi-invariant for $\tau$\/}
if $n(B)=0$ iff $n(\tau B)=0$ and is called {\em conservative for
$\tau$ \/} if $n(W)=0$ for every $\tau$-wandering set $W$.]

Let $f$ be a positive Borel function on $Y$ s.t.\ $\forall\;y$,
the sums $\sum_{k=0}^{\infty} f(\tau^ky)$ and $\sum_{k=1}^{\infty}
f(\tau^{-\,1}y)$ are infinite. Let $T_t,\;t\in\mathbb{R}$ be the
flow $\tau^f$ built under $f$ with base space $(Y,\mathcal{Y})$
and base automorphism $\tau$. It acts on $Y^f=\{(y,t)\;|\; 0\leq t
< f(y),\;y\in Y\}$.

Let $\ell$ denote the Lebesgue measure on $\mathbb{R}$ and let the
measure $n\time \ell$ on $Y\times \mathbb{R}$ be restricted to
Borel subsets of $Y^f$. Let us denote this measure on $Y^f$ by
$m=m_f$.

The flow $T_t,\;t\in\mathbb{R}$, when considered together with the
measure $m$ is called the {\em flow built under $f$ in a measure
theoretic sense}. We call the measure $n$ the {\em base measure}.

Now, \cite{jmmgn}, for any $t\in\mathbb{R}$,\[
\frac{dm_t}{dm}(y,u)=\frac{dn_{_{(t+u)y}}}{dn}(y)\;\;\;\;\mathrm{a.e.m.}\]
where $m_t$ and $n_k$ are the measures $m(T_t(.))$ and
$n(\sigma^k(.))$ respectively and $\frac{dm_t}{dm}$ denotes the
LRN derivative of a quasi-invariant measure \cite{halmos1}. %asdfg

Recall that the flow $T_t,\;t\in\mathbb{R}$, is the flow
$\sigma^f$. Then, as a corollary, we also see that $m$ is
quasi-invariant under the flow $\sigma^f$ iff $n$, the base
measure, is quasi-invariant under $\sigma$. $m$ is invariant under
$\sigma^f$ iff $n$ is invariant under $\sigma$.

Consider now a jointly measurable flow $\tau_t,\;t\in\mathbb{R}$,
on a SBS $(X,\mathcal{B})$ equipped with a probability measure $m$
quasi-invariant under the flow. Let us also assume, for
simplicity, that the flow $T_t,\;t\in\mathbb{R}$, is free. Then,
the map $t\to \tau_t x$ is one-one from $\mathbb{R}$ onto the
orbit $\{\tau_tx\;|\;t\in\mathbb{R}\}$. Thus, a Lebesgue measure
is definable on the orbit simply by transferring the Lebesgue
measure of $\mathbb{R}$ to it. Let us denote by $\ell_x$ this
Lebesgue measure on the orbit of $x$ under the flow.

Then, $m(A)=0$ iff $\ell(\{t\;|\;\tau_tx\in A\})=\ell_x(A)=0$ for
$m$-almost every $x$. [A property which holds for all $x\in X$
except for those $x$ in some $m$-null set is said to hold
$m$-almost everywhere.]

Let $\tau_t,\;t\in\mathbb{R}$, on $(X,\mathcal{B},m)$ and
$T_t,\;t\in\mathbb{R}$, on $(X',\mathcal{B}',m')$ be two
non-singular flows. We say that the two flows are {\em metrically
isomorphic\/} if there exist \begin{description} \item{(i)}
$\tau_t$-invariant $m$-null set $M\in\mathcal{B}$ and
$T_t$-invariant $m'$-null set $M'\in\mathcal{B}'$, \item{(ii)} a
Borel automorphism $\phi$ of $X-M$ onto $X'-M'$
\end{description} s.t.\ $\forall\;t\in\mathbb{R}$, and $x'\in
X'-M'$ we have
\begin{description} \item{(a)} $\phi\circ\tau_t\circ\phi^{-\,1}(x')=T_tx'$,
\item{(b)} $m(\phi^{-\,1}(A'))=0\;\Longleftrightarrow m'(A')=0,\;\forall\;
A'\in\mathcal{B}'$, \item{(c)} in case the flows are measure
preserving we require $m\circ\phi^{-\,1}=m'$ in place of above
(b).\end{description}

Then, as shown in \cite{sgdani}, every non-singular free flow
$\tau_t,\;t\in\mathbb{R}$, on a SPS $(X,\mathcal{B},m)$ is
isomorphic to a flow built under a function in the measure
theoretic sense. The function which implements the isomorphism
preserves null sets.

On the other hand, the basic theorem of Ambrose \cite{ambrose}
states that: every free measure preserving flow on a SPS
$(X,\mathcal{B},m)$ is isomorphic to a flow built under a function
in the measure theoretic sense. The function which implements the
isomorphism preserves the measure. This holds also if $m$ is a
$\sigma$-finite measure \cite{krengel}.

\section{Nomenclature etc.}
We have gone to great lengths in reviewing the basics of the
ergodic theory because of their relevance to the earlier mentioned
significance vis-a-vis P-sets. P-sets are open sets, Borel sets,
of the Einstein space $(\mathbb{B}, d)$ and all the above
considerations are applicable to P-sets.

The question is, of course, of extending and modifying the
relevant of the above point-wise description to suit the P-sets
and, needless to say, of extracting physical results from any such
mathematical analysis.

To this end, we define some nomenclature that will be used in
future relevant works. We note that some of these concepts
coincide with already existing mathematical notions.

Firstly, we reserve the word {\bf particle} to refer to a single
P-set. We shall refer to any collection of P-sets as an {\bf
object} or {\bf objects}. Therefore, an {\em object\/} is a {\em
collection of particles}. In the situation that two or more
objects unite to become a single object, we shall refer to this as
a {\bf merger of objects}. On the other hand, if an object splits
into two or more than two objects, we shall refer to this as a
{\bf splitting of an object}.

Secondly, if a particle splits into two or more particles, we will
call this process a {\bf creation of particles}. On the other
hand, two or more particles unite to become a single particle, we
will call this process an {\bf annihilation of particles}.

To make mathematically more precise the above ideas and also to
conceptually visualize the processes under consideration, the
following will be a useful tool.

Consider two non-empty subsets A and B of a topological space $(X,
\Gamma)$. We call the sets $A$ and $B$ {\em mutually separated\/}
if $A\bigcap B=\emptyset$, {\em ie}, they are mutually disjoint,
and if $A^c\bigcap B=\emptyset$ and $A\bigcap {B}^c=\emptyset$
where ${A}^c$ denotes the closure of the set in question.
Similarly, we define {\em mutually touching non-empty sets A and B
\/} as those sets for which $A\bigcap B=\emptyset$ but
${A}^c\bigcap B\neq \emptyset$ and $A\bigcap {B}^c \neq
\emptyset$. This describes our intuitive notion of touching
particles if the sets under consideration are P-sets.

Then, if two touching P-sets {\em merge\/} under the action of a
Borel automorphism on $\mathbb{B}$, then the boundaries of those
P-sets will merge. If a P-set {\em splits\/} into two or more
P-sets, then the boundary will also split. The above
characterization will be useful to describe such processes.

An object is then some suitable collection of such mutually
touching sets, {\em ie}, it is essentially a region bounded by the
vanishing of the differential of the volume measure (of eq.
(\ref{volume1})) but there are interior points of the region for
which the same differential of the volume measure vanishes, so
such a region is not a P-set.

Various attributes of a physical particle will be Borel measures
on a P-set. Then, using the volume form (of eq. (\ref{volume1})),
we may define appropriate quantities averaged over a P-set.

For example, we may define a suitable a.e. finite-valued, positive
definite, measurable function, the energy density, on a P-set of
$(\mathbb{B}, \mathcal{B},d)$. When this energy density is
integrated wrt the volume measure over a P-set, we may call the
resultant quantity as a mass of the physical particle associated
with that P-set.

We may associate a Dirac $\delta$-function with this mass and then
the ``location'' of a Dirac $\delta$-particle of this mass will be
{\em intrinsically indeterminate\/} over the {\em size\/} of that
P-set.

Such an intrinsic indeterminacy can, thinkably, serve as the
origin of the Heisenberg indeterminacy relations in the continuum
formulation \footnote{Note that Einstein \cite{ein1} regarded the
correctness of Heisenberg's indeterminacy relations as being
``finally demonstrated''. However, as is well known, he differed
from Bohr \cite{ein1} and others on the issue of the ``origin'' of
the indeterminacy relations.}. That this expectation is indeed
true or not will be demonstrated in a later work. Such a
demonstration will, obviously, require us to sharpen
(mathematically) the notion of {\em motion\/} of a particle within
the present formulation.

Now, while considering the ``motion'' of this ``particle with
mass'', we can, of course, have such measures invariant. We will
therefore be considering measure preserving automorphisms of the
base space $\mathbb{B}$ (leaving aside the issue of the effect of
the automorphism on other P-sets).

Just as we may consider Borel automorphisms of the base space
$\mathbb{B}$ that keep a P-set invariant, we may also consider
that an object, a collection of P-sets, remains invariant under
the automorphism of the base space $\mathbb{B}$.

In the analysis of such situations, we may then consider a {\em
countably finite or infinite chain\/} of mutually touching
non-empty sets as a function $f: \mathbb{N}\to \mathbb{B}$ s.t.\
$\left\{ A_0, A_1, ...\right\}$ with $A_i$ and $A_{i+1}$ being
mutually touching sets. (The set of natural numbers may be
replaced by any directed set here.) Such chains may also remain
invariant under the automorphisms of the base space.

At this place, we therefore note that the group of Borel
automorphisms of the Einstein space $(\mathbb{B}, \mathcal{B}, d)$
is sufficiently large so as to permit such possibilities as above.

\section{Concluding remarks}

Now, our ultimate goal is to describe, in precise mathematical
terms, different processes of merger of objects, splitting of an
object, creation of particles, annihilation of particles in the
base or the physical space $\mathbb{B}$.

Furthermore, apart from the above processes, we also want to
describe motions of physical objects and other phenomena using the
same formalism. This is mainly because the current formalism is
supposed to be incorporating the totality of all the possible
fields in Nature.

At this point, we then note that in the physical world, we measure
distances in terms of a certain {\em unit of distance}. Distance
between two given objects is always an integral multiple of the
basic unit of distance. We cannot {\em measure\/} any fraction of
the basic unit of distance unless, of course, we have a physical
object smaller in size than the chosen basic unit of the distance.

For example, let us measure distances in terms of a basic unit -
the centimeter. Then, distance between two physical objects is
always an integral multiple of a centimeter. Unless we find a
physical object smaller than a centimeter, we will be unable to
``measure'' distances smaller than a centimeter. This is the issue
here.

Now, this unit of distance can, of course, be chosen to be a P-set
or a suitable collection of P-sets, {\em ie}, an object. The sizes
of P-sets and objects are mathematically well-defined in the
present formalism. The metric of the Einstein space $(\mathbb{B},
\mathcal{B}, d)$ allows us the precise mathematical description of
these conceptions.

Then, given a P-set or an object, we may construct a finite chain
of such objects. Then, the distance between two objects can also
be measured to be an integral multiple of the ``size of a chosen
finite chain of objects''. This is easily describable within the
present formulation.

Clearly, to measure distances smaller than the chosen size of
object, we will need another object with ``size smaller than that
of the first object''. Such P-sets and objects always exist is
what has emerged in the present formalism.

We have then seen that relevant results from the ergodic theory
imply the existence of a suitable Polish topology on the Einstein
Borel space $(\mathbb{B}, \mathcal{B})$ whose certain open sets
form the class of all P-sets of the metric space $(\mathbb{B},d)$.

It is then intuitively clear that, using the P-sets or objects, we
can always implement the aforementioned construction.

This suggests an appropriate ``distance function'' on the family
of all P-sets/objects of the base space $\mathbb{B}$. Intuitively,
it is this {\em physical distance\/} that changes with {\em time
parameter\/} and is the motion in the physical world. This then
constitutes the {\em dynamics in the physical world}.

After all, motion of one body is to be described relative to
another body. Of course, we can also consider that two or more
bodies do not move relative to each other.

Then, the Borel automorphism of the Einstein space may result into
change in the physical distance resulting into relative motion of
bodies. On the other hand, an automorphism that keeps invariant a
chain of P-sets separating two P-sets can describe the situation
of two or more relatively stationary objects.

It is therefore {\em crucial to distinguish\/} between the {\em
mathematical metric\/} (\ref{3d-metric-gen}) and the above
mentioned ``metric on the class of all P-sets'' which we will call
a {\em physical metric}.

We also note here that we can consider countable collections of
P-sets and, hence, of objects as defined above. Then, the fact
that we can ``count objects'' in the present formalism also agrees
well with our general experience.

Physical objects, each one being an appropriate collection of
particles, are ``countable'' in Nature. For example, we can count
the number of chairs, tables, persons in a conference hall. Such
physical objects are simply collections of objects. Any
theoretical formulation must also be able to describe them as
such. This is then the obvious motivation behind our adopting
various earlier definitions and nomenclatures which also possess,
already existing, mathematical analogues.

The physical metric mentioned above then provides the mathematical
notion of the physical distance between the objects. Effects of
the Borel automorphisms of the base space on the physical metric
are then to be looked upon as resulting physical motions of
physical bodies and various physical phenomena as manifestations
of relevant properties of such automorphisms \footnote{ Therefore,
a joint manifestation of Borel automorphisms of the Einstein Borel
space $(\mathbb{B}, \mathcal{B})$ and the association of a Dirac
$\delta$-particle with a P-set is behind Heisenberg's
indeterminacy relations in the present continuum formulation.}.

Essentially, our point of view here is therefore that the {\em
physical dynamics\/} arises as a result of the Borel automorphisms
of the Einstein Borel space $(\mathbb{B}, \mathcal{B})$ and their
effects on some suitable physical distance function over the class
of P-sets of the Einstein metric space $(\mathbb{B}, d)$.

Then, since we treat physical attributes of a particle as suitable
measures, it is also the contention here that the group of Borel
automorphisms of the Einstein space $(\mathbb{B}, \mathcal{B}, d)$
is sufficiently large to encompass the observed physical
phenomena. It is in this broad sense that the present formalism is
to be a theory of everything.

Now, one is also struck by the fact that, in the usual formalism
of General Relativity, the measuring apparatuses, {\it e.g.}, rods
and clocks, are postulated in a manner that isolates them from the
physical phenomena that are being described. For example, the
measuring apparatuses in a given spacetime are postulated to be
independent of the presence of electromagnetic or other types of
fields imposed on that spacetime.

Strictly speaking, measuring apparatuses will have to be treated
at par with every other thing that the theory treats, and not be
theoretically self-sufficient entities that remain isolated from
other physical entities forever.

Then, it is worth pointing out here that the present formalism
represents measuring apparatuses in precisely this desired manner.
The effects of (Borel) automorphisms of the Einstein space
$(\mathbb{B}, \mathcal{B}, d)$ on the measuring apparatuses are
also accountable in it. Moreover, their treatment \footnote{ A
measuring rod is a suitable chain of P-sets or objects. A
measuring clock is a P-set or an object undergoing periodic
motion. That their treatment in the present formalism is the same
as that of everything else is self-evident. \\ The comprehension
of such obvious ``measuring apparatuses'' will require us to work
out precisely the corresponding Borel automorphisms and their
effects on the P-sets or objects under consideration. This type of
treatment will be presented in a later work.} is also at par with
the treatment of every other thing that the formalism treats.

The present effort constitutes an attempt of developing a theory
of the physical universe based on the continuum picture. That it
possesses the required simplicity \footnote{Recall here a
statement by Einstein that \cite{ein1}: A theory is the more
impressive the greater the simplicity of its premises is, the more
different kinds of things it relates, and the more extended is its
area of applicability.} while simultaneously encompassing number
of physical phenomena needs to be stressed then. Again, this is
due to the fact that the present formalism is to incorporate all
the fields of Nature in it.

We then note that the Borel automorphism of the Einstein space
$(\mathbb{B}, \mathcal{B}, d)$ {\em cannot\/} lead to the
formation of a {\em singularity\/} in any conceivable manner
because any such automorphism is a one-to-one and onto map that is
also measurable with its inverse also being a measurable map.
Consequently, any naked singularities do not arise here as the
singularities themselves do not arise.

The question of whether event horizons or black holes can arise
within the present formalism is a more complicated one to settle
than that of naked singularities.

However, the Borel automorphisms of the Einstein space
$(\mathbb{B}, \mathcal{B}, d)$ form a group. Then, we can always
traverse \footnote{Let a Borel automorphism of the Einstein space
$(\mathbb{B}, \mathcal{B}, d)$ result into a entry into the given
2-surface from one direction. The inverse of that Borel
automorphism (resulting into entry in the other direction out of
that 2-surface) exists to reverse the first entry is the point
here.} or cross any given 2-surface both ways. This suggests that
a one-way membrane may not arise in the present formalism. It
therefore appears that even black holes may not arise in the
present formalism.

Above ideas, of physical (as well as mathematical) nature, are
mainly intuitive and these ideas will have to be formulated in
precise mathematical terms. It is this task that we devote to in a
series of papers following this one.

In conclusion, we however note the following. A particle as a
concept has existence only at a single spatial location. On the
other hand, a continuum has simultaneous existence at more than
one, continuous, locations. These are primarily antipodal
conceptions.

Newtonian theory is an attempt to understand the happenings of the
physical world on the primary basis of the conception of a
point-particle. Newton's mechanical world-view codifies all that
can be consistently achieved using it.

On the other hand, General Relativity attempts to dispense with
this notion of a particle and replaces it with that of an extended
particle, the continuum of energy in space.

General Relativity is an attempt to grasp the happenings of the
physical world on the basis of the continuum hypothesis. In a
definite sense, it is an extension,  to include gravity, of the
field conception of Faraday and Maxwell in the form of
electrodynamics. But, since the laws of electrodynamics of Maxwell
and Faraday were linear laws, this extension to gravity required
intrinsically non-linear laws. These non-linear laws are the
Einstein field equations relating the geometric quantities to the
matter quantities.

In the present paper, we argued as to why the continuum
description of General Relativity requires the use of a specific
spacetime geometry. We then argued as to why we need to restrict
to only the 3-dimensional manifold of the Einstein pseudo-metric
(\ref{3d-metric-gen}). This provided us the required notion of
extended bodies, bodies as concentrated form of energy in space.

We then showed that the base space of (\ref{3d-metric-gen}) is a
Standard Borel Space. The description of the motion of an
extended-particle is then permissible along the lines proposed
herein.

Whether this description agrees with various experimental or
observational results then remains to be investigated and is some
issue.

\begin{acknowledgments}
I am grateful to Pankaj Joshi, Ravi Saraykar, Girish Sahasrabudhe,
Shreerang Thergaonkar,  Vivek Wagh, M. G. Nadkarni and S. G. Dani
for discussions and helpful comments that led to clarification of
my own ideas. Of course, any errors of understanding are entirely
my responsibility.
\end{acknowledgments}

 \medskip
\appendix

\section*{Glossary of notations}
\begin{description} \item{SBS} \hfill Standard Borel Space
\item{SMS}  \hfill Standard Measure Space \item{SPS}  \hfill Standard
Probability Space \item{wrt}  \hfill  with respect to \item{s.t.}
\hfill such that \item{a.e.} \hfill almost everywhere
\item{$\mu$-a.e.}  \hfill modulo a set of $\mu$-measure zero
\item{fimp}  \hfill for infinitely many positive \item{fimn}  \hfill
for infinitely many negative \item{{\it s}$_{_T}(A)$}  \hfill
$=\,\bigcup_{k=-\infty}^{\infty} T^kA$ : \hspace{.1in} {\em
$T$-saturation of $A$}
\item{B-shift} \hfill  Bernoulli shift \item{K-shift} \hfill Kolmogorov shift
\item{DAM}  \hfill Diadic Adding Machine or Odometer \item{$\mathfrak{P}(X)$}
\hfill Power set of a given set $X$ \item{$A\triangle B$} \hfill
Symmetric union $=(A-B)\bigcup(B-A)$
\item{$\mathcal{B}$ or $\mathcal{B}_{_X}$} \hfill Borel $\sigma$-algebra of
$X$ \item{$\mathcal{N}$} \hfill $\sigma$-ideal of subsets of a
given set $X$
\item{$A\sim B$} \hfill equivalence of sets $A$ and $B$ by \\
\phantom{m} \hfill countable decomposition
\item{$A\sim B \,(\mathrm{mod} \,\mathcal{N})$} \hfill equivalence
of sets $A$ and $B$ by \\ \phantom{m}  \hfill countable
decomposition $(\mathrm{mod} \,\mathcal{N})$ \smallskip
\item{$\mathcal{W}_{_T}$} \hfill $\sigma$-ideal generated by all
$T$-wandering \\ \phantom{m} \hfill sets in $\mathcal{B}$ or the
Shelah-Weiss ideal
\item{$\mathcal{H}$} \hfill the Hopf ideal \item{$A\prec\prec
B$} \hfill if $\exists$ a measurable subset $C\subseteq B$ s.t.\\
\phantom{m} \hfill $A \sim C$ and {\it s}$_{_T}(B-C)\;=\;${\it
s}$_{_T}B$
\item{$A\prec\prec B\;(\mathrm{mod}\; \mathcal{N})$} \hfill if
$\exists$ a set $N\in\mathcal{N}$ s.t.\\ \phantom{m}\hfill
$A\triangle N \prec\prec B\triangle N$ \item{$A$ is {\em
$T$-compressible}} \hfill if $A\prec\prec A$
\item{$T_1 \prec T_2$} \hfill if $T_1$ is isomorphic to $(T_1)_{_A}$
for some \\ \phantom{m} \hfill $A\in\mathcal{B}$ with
$\bigcup_{k=0}^{\infty}T^k_1A=X$, that is, \\ \phantom{m} \hfill
if automorphism $T_1$ is a {\em derivative} \\ \phantom{m} \hfill
of $T_2$ or if $T_2$ is a {\em primitive} of $T_1$
\item{$T_1 \kakueq T_2$} \hfill Kakutani
equivalent automorphisms
\end{description}

\end{document}